\documentclass[referee]{aa}
\usepackage[varg]{txfonts}

\usepackage{graphicx}
\usepackage{natbib}
\usepackage{color}
\usepackage{subfigure}
\usepackage[switch]{lineno} 
\usepackage{xspace}
\usepackage{microtype}
\usepackage{amssymb}
\def\link_col{blue}
\usepackage{xspace}
\usepackage{url}
\usepackage{wasysym}

\def\grays{$\gamma$-rays\xspace}
\def\gray{$\gamma$-ray\xspace}

\begin{document}

\title{On the shape of the gamma-ray spectrum  around the ''$\pi^0$-bump'' }
\titlerunning{pion-decay bump}
\author{Rui-zhi Yang\inst{1}
\and Ervin Kafexhiu\inst{ 1}
\and Felix Aharonian\inst{2,1, 3}}
\institute{Max-Planck-Institut f{\"u}r Kernphysik, P.O. Box 103980, 69029 Heidelberg, Germany.
\and Dublin Institute for Advanced Studies, 31 Fitzwilliam Place, Dublin 2, Ireland.
\and MEPHI, Kashirskoe shosse 31, 115409 Moscow, Russia
}%

\date{Received:  / Accepted: } 
\abstract {The "pion-decay" bump  is a distinct signature of the differential 
energy spectrum of
$\gamma$-rays between 100~MeV and 1~GeV produced in hadronic interactions of accelerated particles (cosmic rays) with the ambient gas.  We use the recent parametrisations of  relevant 
cross-sections  to study the formation of the "pion-decay" bump. 
The $\gamma$-ray spectrum below the maximum of this spectral feature can be distorted because of contributions of
additional radiation components, in particular, due to the bremsstrahlung of secondary electrons and positrons, the products of decays of  $\pi^\pm$-mesons, accompanying the $\pi^0$-production.  At energies below 100 MeV, a non-negligible fraction  of $\gamma$-ray flux could originate  from interactions of sub-relativistic heavy ions.  We study the impact of these radiation channels  on the formation 
of the overall  $\gamma$-ray spectrum   based on a time-dependent treatment of  evolution of  energy distributions of  the primary and secondary particles in the  $\gamma$-ray production region.}


\maketitle
\section{Introduction}
Cosmic Rays (CRs)  produce high energy \grays via hadronic  interactions  with the ambient gas. The neutral $\pi^0$ mesons appearing  in these collisions promptly decay into two $\gamma$-ray photons each having an energy of $m_{\pi^0}/2 = 67.5~ \rm MeV$ (in the rest frame of the  pion). 
The \grays are symmetrically distributed around $67.5$ MeV in the log energy scale.  The broad-band gamma-ray spectra resulting from these processes have been studied in  various astrophysical environments such as Solar Flares,  
Interstellar Medium, SNRs, Molecular Clouds,  Galaxy Clusters, {\it etc.} \citep{murphy87, pfrommer04, ohira12}.  The spectral energy distribution (SED) of this radiation, $E^2 \frac{dN}{dE}$,  has a distinct  bell-type feature (``$\pi^0$-decay bump'') between $100~\rm MeV$ and  a few GeV\citep{Stecker71} .  The shape and the position of the maximum 
of this feature  depends  on the spectral index $\alpha$ of the energy
distribution of  parent protons and nuclei. In particular,  the bump disappears 
completely  for  $\alpha < 2$.  The $\pi^0$-decay bump  claimed to be detected 
by the  AGILE  and  {\it Fermi}  LAT  collaborations towards several mid-aged supernova remnants (SNRs) \citep{agile_pion,fermi_pion, tavani10, giuliani11},  generally  is interpreted  as an evidence of acceleration of  cosmic-ray protons and nuclei  in SNRs.   This spectral feature,  however,  does not  appear in a "pure" form. 
A significant fraction of the \gray flux in this energy range can be contributed by the 
bremsstrahlung of primary (directly accelerated) electrons. This could happen, in particular, in the environments with the CR electron to proton ratio $e/p \geq 0.1$ \citep[see, e.g., ][]{aharonian04}. Moreover, even when the acceleration is  strongly dominated by the hadronic component of CRs, 
the production of secondary electrons and positrons and their consequent radiation in the $\gamma$-ray band  is unavoidable. 
The contribution of these  radiation channels depends on the density of the ambient gas, $n$,  and the confinement time of CRs, $T$. It achieves  its maximum (the "saturation") when 
the product $n \times T \simeq  5\times10^{15} ~\rm s/cm^3$ (see below). Depending on the spectrum of  CR nuclei and their composition, a non-negligible contribution to $\gamma$-rays below 100 MeV 
is  expected also from the reactions induced by subrelativistic nuclei.  In this paper, we 
study the  relative contributions  of  these radiation channels to the formation of 
the overall $\gamma$-ray  spectrum in the  region of the $\pi^0$-decay bump.

\section{Time-dependent energy distribution of particles}
The evolution of relativistic particles in a given volume is described by the kinetic equation \citep[see e.g., ][]{ginzburg64}
\begin{equation}
\frac{\partial N}{\partial t}=\frac{\partial}{\partial E}\left(P\:N \right)-\frac{N}{\tau_{esc}}+Q\:,
\end{equation}
where $P=P(E)=-\frac{dE}{dt}$ is the energy loss rate and $\tau_{\rm esc}$ is the characteristic escape time.   For simplicity, in the following discussion we will  neglect the particle escape from the $\gamma$-ray production region.  The escape of CRs from their acceleration sites is far from being understood. It strongly depends on the diffusion coefficient which in the $\gamma$-ray production region is a highly unknown parameter.  Generally,  for low-energies, this could be a good approximation
given the slow diffusion near the CR sources and the age of the accelerator. For example in a source of size $R\sim 10$~pc, the escape time is estimated $t_{\rm esc} \sim R^2/ 3 D \geq 10^{5}$~yr assuming that the diffusion coefficient at subrelativistic energies does not exceed $10^{26} \rm cm^2/s$.  In any case, the ignorance of the particle escape gives  upper limits on the contribution of secondary electrons to the overall $\gamma$-ray flux.

For continuous injection  $Q(E,t)=Q(E)$, the solution of the kinetic equation becomes
\begin{equation}\label{Nsolution}
N(E,t)=\frac{1}{P(E)} \int^{E_{0}}_{E}Q(E)dE\;,
\end{equation}
where $E_{0}$ is found by solving the characteristic equation for the given epoch $t$, $t=\int^{E_{0}}_{E}
\frac{dE}{P(E)}$. 

\begin{figure*}
\centering
\subfigure[][$\Gamma=-2$ in momentum]{
\includegraphics[scale=0.65]{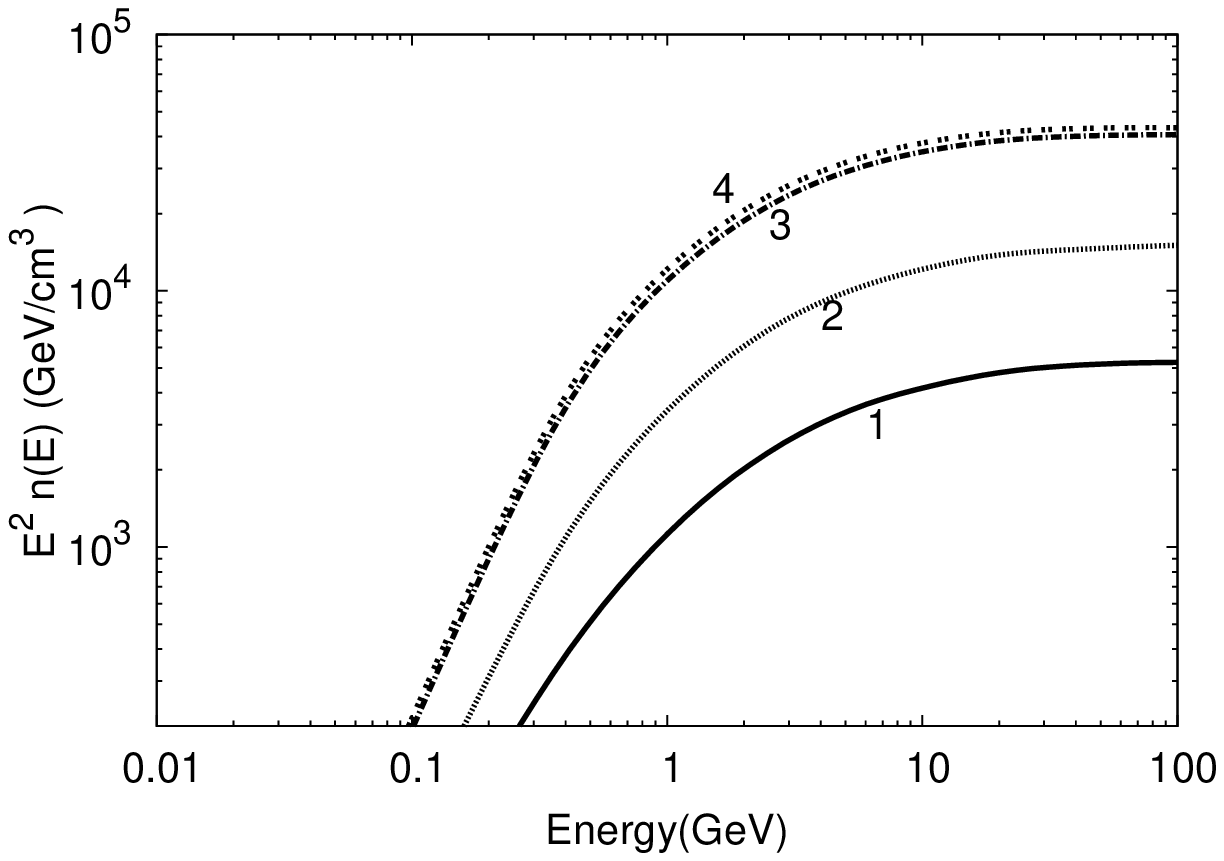}
}
\subfigure[][$\Gamma=-2.85$ in momentum]{
\includegraphics[scale=0.65]{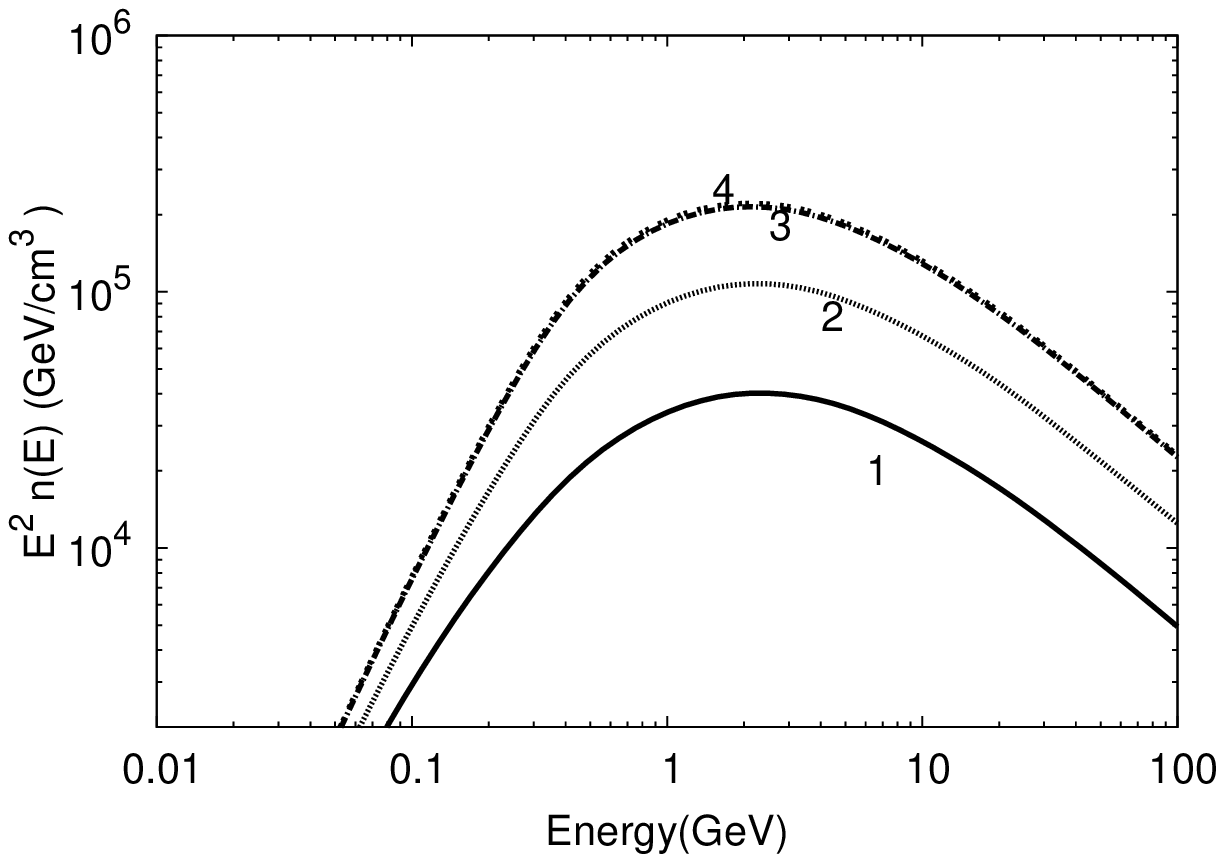}
}

\caption{Energy distributions of CR protons at different epochs characterised by the  
parameter $n\times T$ values (in unit $\rm cm^{-3} s$):~ $3 \times 10^{14}$  (curve 1),  $10^{15}$ (curve 2),  $5 \times 10^{15}$ (curve 3), $10^{16}$ (curve 4).  The left panel is for the proton injection spectrum with an index of 2, and the right panel for the index 2.85 (in momentum). The normalisations are arbitrary. }
\label{fig:prosed}
\end{figure*}

For protons, the dominant cooling mechanisms are the nuclear interactions and the ionisation losses. The nuclear interactions dominate above 1 GeV and can be characterized  as $P_{nuc}=E/\tau_{pp}$, where E is the proton energy and $\tau_{pp}$ is the nuclear interaction loss timescale determined by the total inelastic cross section $\sigma_{pp}$ and the inelasticity $\kappa$. For proton energies above $1~\rm GeV$,  $\sigma_{pp} ~\sim 30~\rm mb$ and $\kappa \sim 0.45$; both only slightly depend on energy. Then $\tau_{pp}$ can be expressed as $\tau_{pp} = (nc\kappa \sigma_{pp})^{-1}= 6\times 10^5\frac{1~\rm cm^{-3}}{n} yr$.  The ionisation losses dominate in the low energy domain. The energy  loss rate $P_{ion}$  is proportional to the ambient gas density. At high energies $P_{ion}$ is energy independent, whereas, at low energies, between $1~\rm MeV$ and $1~\rm GeV$, the $P_{ion}$ scales as $1/\beta$, where $\beta = v/c$. Convenient  analytical presentations  for  ionisation losses can be found e.g.  in \citet{gould72}.
The injection spectra are assumed to be power laws in  momentum with indexes 2.0 and 2.85, that is, 
\begin{equation}
 Q(E) \sim \frac{N(p_0)}{\beta c} \left(\frac{p}{p_0}\right)^{-\Gamma}, 
 \label{eq:time}
 \end{equation}
 where $E$ is the protons total energy, $p$ is the proton momentum, and 
 $\beta$ is  the proton velocity in units of c.
 Fig.\ref{fig:prosed} shows the derived proton distributions at different epochs. 
Note that for 
\begin{equation}
 n \times T\sim5\times10^{15} ~\rm s/cm^3,
 \label{eq:time}
 \end{equation}
 the evolution of protons saturates,  thus  the source of age $T$ operates as a calorimeter. 
 Then, the established ("steady-state") density of CRs  can be estimated as $N \sim Q ~t_{cool}$. Note that for both nuclear and ionisation losses, the loss timescales are inversely proportional to the ambient gas density, therefore 
 $ n \times T$ is the only quantity to determine whether the system saturates or not.  For the dense clouds with the gas number density of $100-1000~\rm cm^{-3}$,  the characteristic time $T$ for saturation is about $10^5 -10^6~\rm yrs$. At low energies  a partial saturation, depending on the diffusion coefficient,  can be achieved in these dense structure. On the other hand, in ISM, where $n\sim 1~\rm cm^{-3}$,  the required time $T$ for saturation is about $10^8\rm yrs$, which is much longer than the confinement time of the CRs inside the Galaxy.  In contrary,  in the giant halo surrounding the Galaxy, the gas density is much lower, however  the confinement time can be much larger. This may set more favoured conditions for saturation in the halo rather than in the galactic disk \citep[see e.g., ][]{taylor14}.  

The combination of ionisation and nuclear loses results in a formation of a break in the 
particle spectrum during its evolution. In particular, at the stage of saturation,  
for the  power law injection spectrum $Q \propto E^{-\Gamma}$ and  the energy lose rate depending on
the energy as 
$P={\rm d}E/{\rm d}t \propto E^{\alpha}$, the power-law index of the resulted proton spectrum  becomes
$\Gamma'=\Gamma+\alpha-1$ (see Eq.(3)).  
At high energies, when the nuclear interactions dominate, 
the energy loss rate is approximately proportional to the proton energy, 
$P_{\rm nuc} \sim E $, thus the nuclear interactions practically do not change the injection spectrum of protons. This is seen in  Fig.\ref{fig:prosed}, where  at energies above 1 GeV the saturated proton spectrum keeps the 
initial power-law index. At energies below 1 GeV, more important are ionisation losses for which  $\alpha  \approx 0$.  Consequently,  $\Gamma'  \approx \Gamma-1$.  

\begin{figure*}
\centering
\subfigure[][$\Gamma=-2$ in momentum]{
\includegraphics[scale=0.65]{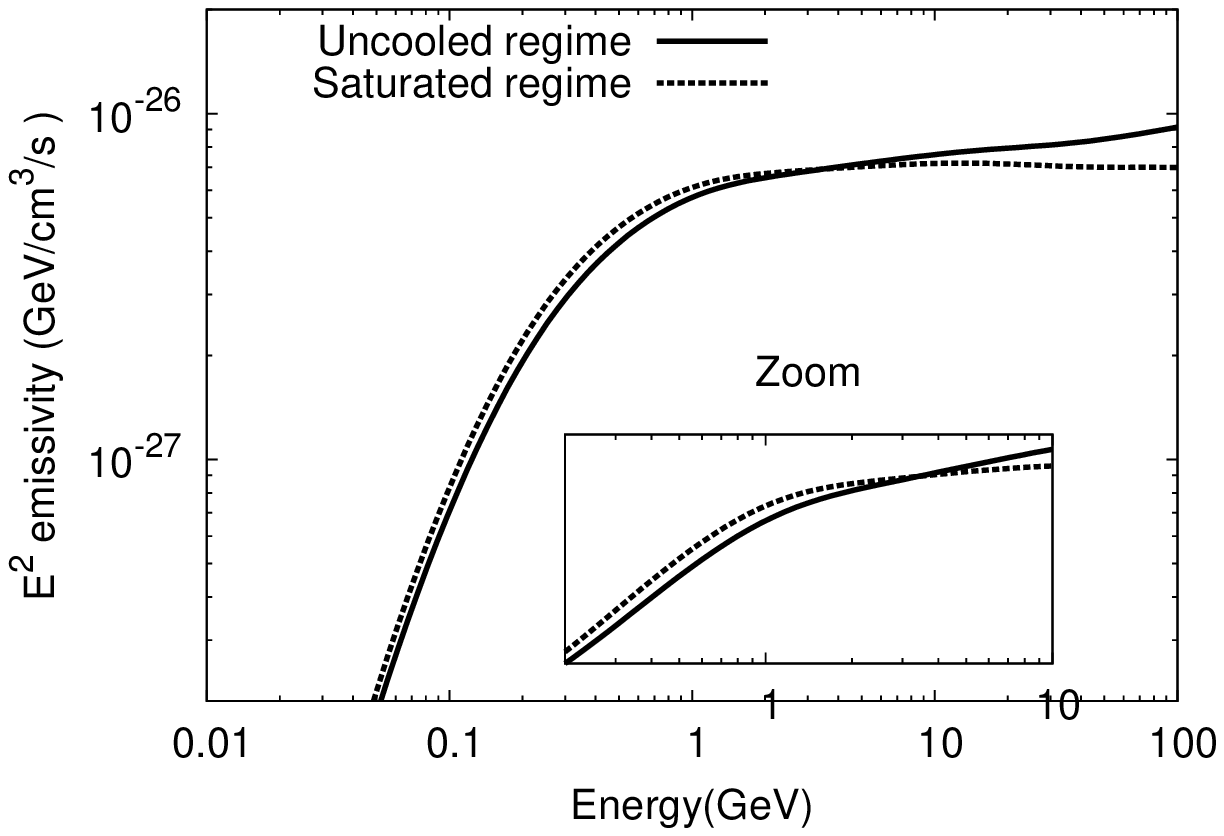}
}
\subfigure[][$\Gamma=-2.85$ in momentum]{
\includegraphics[scale=0.65]{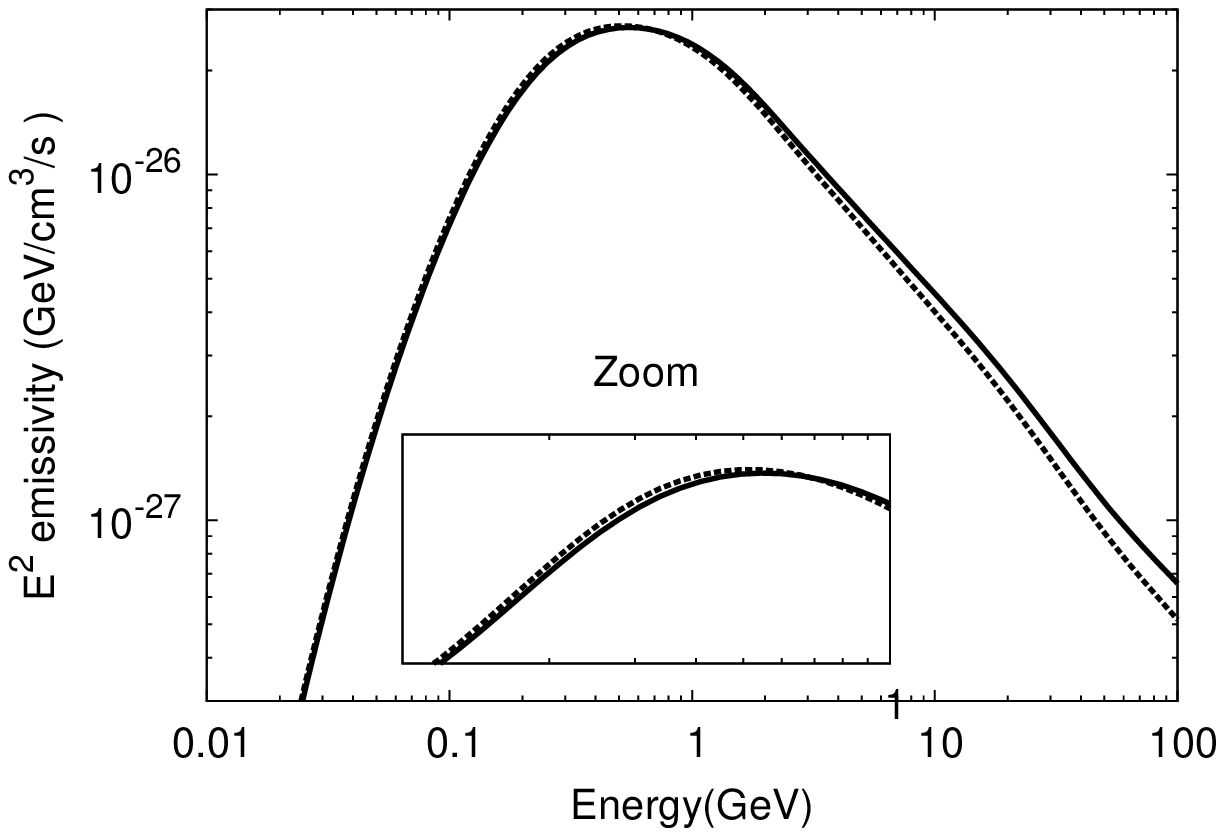}
}
\caption{Gamma ray emissivities (in the form of  spectral energy distribution - SED) for the uncooled and saturated CR protons. The number density of the ambient gas is taken $n=1 \ \rm cm^{-3}$. The proton distributions are normalised in a way that the energy density above $T=1$~GeV is equal to $1~\rm eV/cm^3$. The inserted figures are  the zoomed SEDs  in the maximum region between 0.1 and 10 GeV.}
\label{fig:emi}
\end{figure*}

The differences  in  the uncooled and saturated proton spectra at low energies are transferred to the  spectra of $\gamma$-rays. The impact of this effect, however,  is  rather small, as it  can be seen 
in Fig.\ref{fig:emi}. This is explained by the close location of the cooling break 
in the proton spectrum around 1~GeV to the 
kinematic threshold of production of $\pi^0$-mesons at  300 MeV.  Note that  the difference between the 
the $\gamma$-ray spectra formed in  the protons uncooled and saturation regimes,  is small also at high energies, $E \geq 10$~GeV (see Fig.\ref{fig:emi}), but for a different reason.  
In the uncooled regime,  the  \gray emission is harder than that of the parent protons due to the increase of the total $\pi^0$-decay  cross-section with energy. In saturated regime, however, the proton spectrum should be slightly softer than the uncooled (injected) proton spectrum. This softening  is caused by the same energy dependence of the cooling rate, which is proportional to the total inelastic p-p cross section. Such a softening compensates, to a large extent,  the spectral hardening in the \gray production.

Although Fig.\ref{fig:emi} demonstrates  a rather weak  time-evolution of the  
energy spectra of  $\pi^0$-decay  $\gamma$-rays, we should note that it is
true only in the case of an effective confinement of protons in the $\gamma$-ray production region,
i.e. when $t_{\rm pp} \leq t_{\rm esc}$.  Otherwise, the energy-dependent escape would lead to a strong evolution of  the proton spectrum.  And, apparently, this will be reflected in the 
spectra of secondary $\gamma$-rays  

\begin{figure*}
\centering
\subfigure[][$\Gamma=-2$  in momentum]{
\includegraphics[scale=0.65]{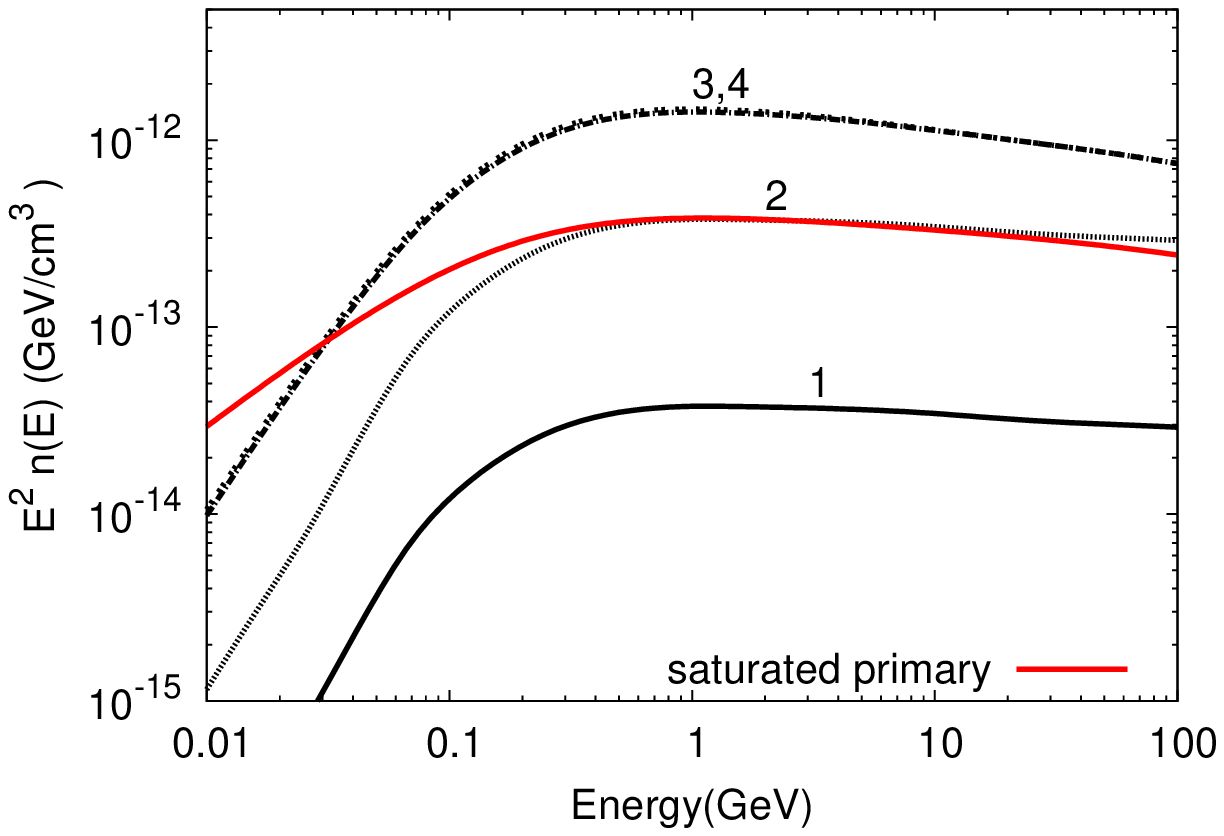}
}
\subfigure[][$\Gamma=-2.85$  in momentum]{
\includegraphics[scale=0.65]{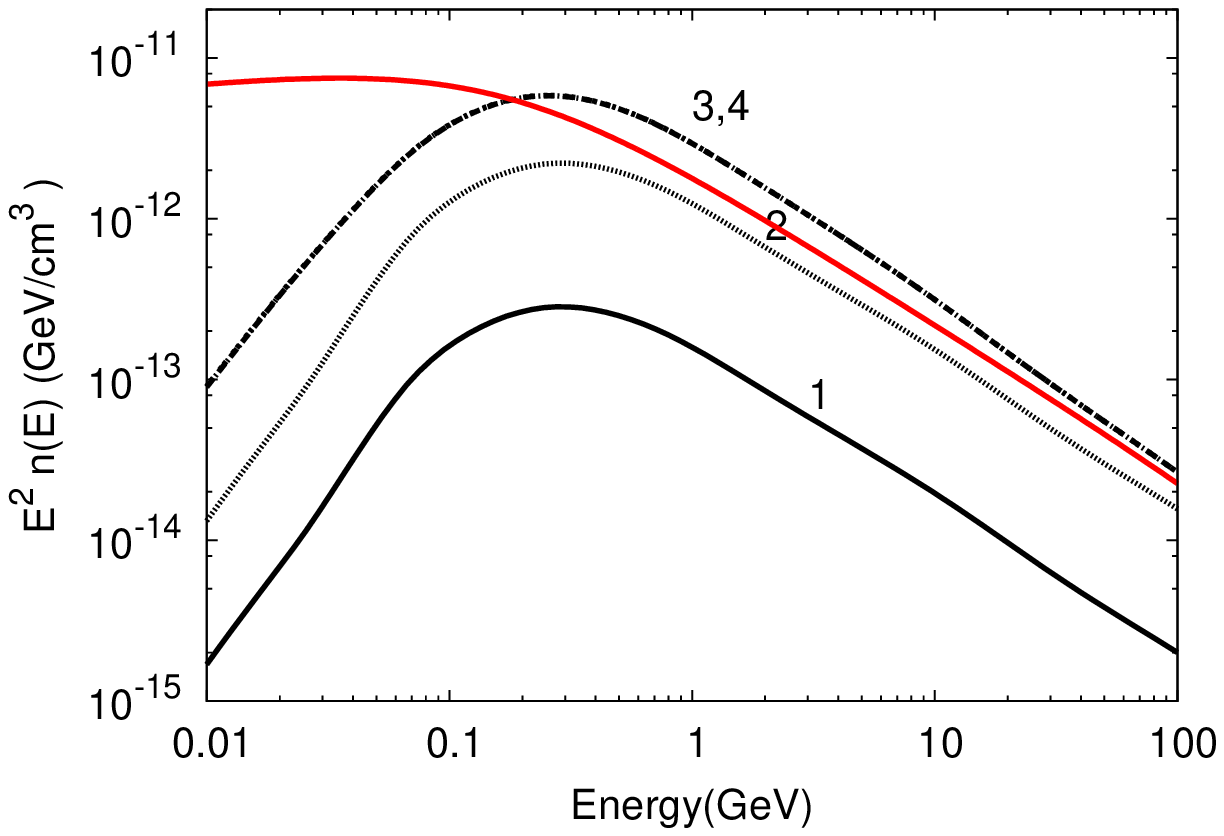}
\label{OrionB:4}
}
\caption{The time evolution of distributions of the secondary electrons (black curves) calculated for 
 different values of $n\times T$ (in unit $cm^{-3} s$):  $10^{13}$ (curve 1), $10^{14}$ (curve 2), 
  $10^{15}$ (curve 3),  and  $3 \times10^{15}$ (curve 4). The parent proton densities are normalised such that the energy density above 1~GeV is $1~\rm eV/cm^3$.  The saturated spectra for primary electrons (red curves) are also shown. The primary electrons are assumed to have the same spectrum as the parent protons, and the e/p ratio of 0.01 at injection. }
\label{fig:ele_sed}
\end{figure*}

The steady-state distributions of electrons and positrons are also described by Eq.(2).  However, their cooling is caused, in addition to the ionisation losses, by the synchrotron, bremsstrahlung and the Inverse Compton (IC) radiation channels.  Below,   we fix  the ambient gas density and the magnetic field to 
the values   $100 ~\rm cm^{-3}$ and $10~\rm \mu G$, respectively. The Interstellar radiation fields are assumed to  comprise three components: (1)  the 2.7 K CMB with an  energy density of $0.24 ~\rm eV/cm^3 $,  (2) the optical/UV field modelled as a grey body component with  an energy density of $2 ~\rm eV/cm^3 $ and temperature of $5000~\rm K$  and (3) the IR component which is modelled  as a grey body component with  an energy density of $1 ~\rm eV/cm^3 $ and temperature of $100~\rm K$. 
The  time evolution of the spectrum of secondary electrons  is shown in Fig. \ref{fig:ele_sed}  assuming  a constant injection rate of  parent protons with a  spectrum normalised to the energy density of $1~\rm eV/cm^3$ above $1~\rm GeV$. The electron spectrum saturates when $n\times T$ exceeds $10^{15} ~\rm s/cm^3$.   

For the chosen parameters,   the energy losses of electrons in the energy interval between several 100 MeV to several 100 GeV,  are dominated by bremsstrahlung. Since  the energy-loss rate of  bremsstrahlung is  nearly energy-independent,   the saturated secondary electrons and positrons have the same  spectral shape as the parent protons. Below several hundred MeV, the ionisation losses start to dominate which results in a low energy break in the electron spectrum. We also plot the saturated spectrum of primary electrons, the injection spectrum of which is assumed to be a power law ( in momentum ). The saturated primary electron spectrum also has a low energy break, but shallower than that in the spectrum of secondary electrons; the break in the primary spectrum is caused by the ionisation losses, while the break in the spectra of secondaries is caused by both the ionisation losses and the injection spectrum from the charged pion decays. 

\section{$\gamma$-ray production in hadronic processes}

\subsection{Cross sections}

The calculations  of \gray  spectrum near the pion-decay bump require good 
knowledge of  the $\pi^0$ production cross-sections at  the proton-proton, proton-nucleus, and nucleus-nucleus interactions.  The theoretical predictions of these cross-sections are limited by the presence of non-perturbative processes. Therefore different phenomenological treatments  have been 
discussed  in the literature. At high energies, a
convenient and comprehensive parametrisations of the cross sections for the production of \grays, electrons and neutrinos have been proposed by \citet{kelner06}. This parametrization, however,  is not designed for precise  calculations  at low energies,  in particular  near the kinematic threshold of $\pi$-production. Recently, a new parametrization  for production of $\gamma$-rays in a broader energy range of $pp$ interactions, from the kinematic threshold to PeV energies,  has been proposed by  \cite{kafexhiu14}.  Most importantly, this parametrization allows accurate description of  experimental data in the most relevant energies for the formation of the bump region, below $T_p<2$~GeV.   At such low energies,  the previous parametrizations  \citep{dermer86,Kamae2006} show  significant deviations from the experimental data \citep{kafexhiu14}.  At higher energies, the parametrization of \citep[see, e.g., ][]{kafexhiu14} offers a choice of switching between different hadronic models to account for the uncertainties in the experimental data. 
This parametrisation  smoothly connects the low and high energy regions.

\begin{figure}
\centering
\includegraphics[scale=0.54]{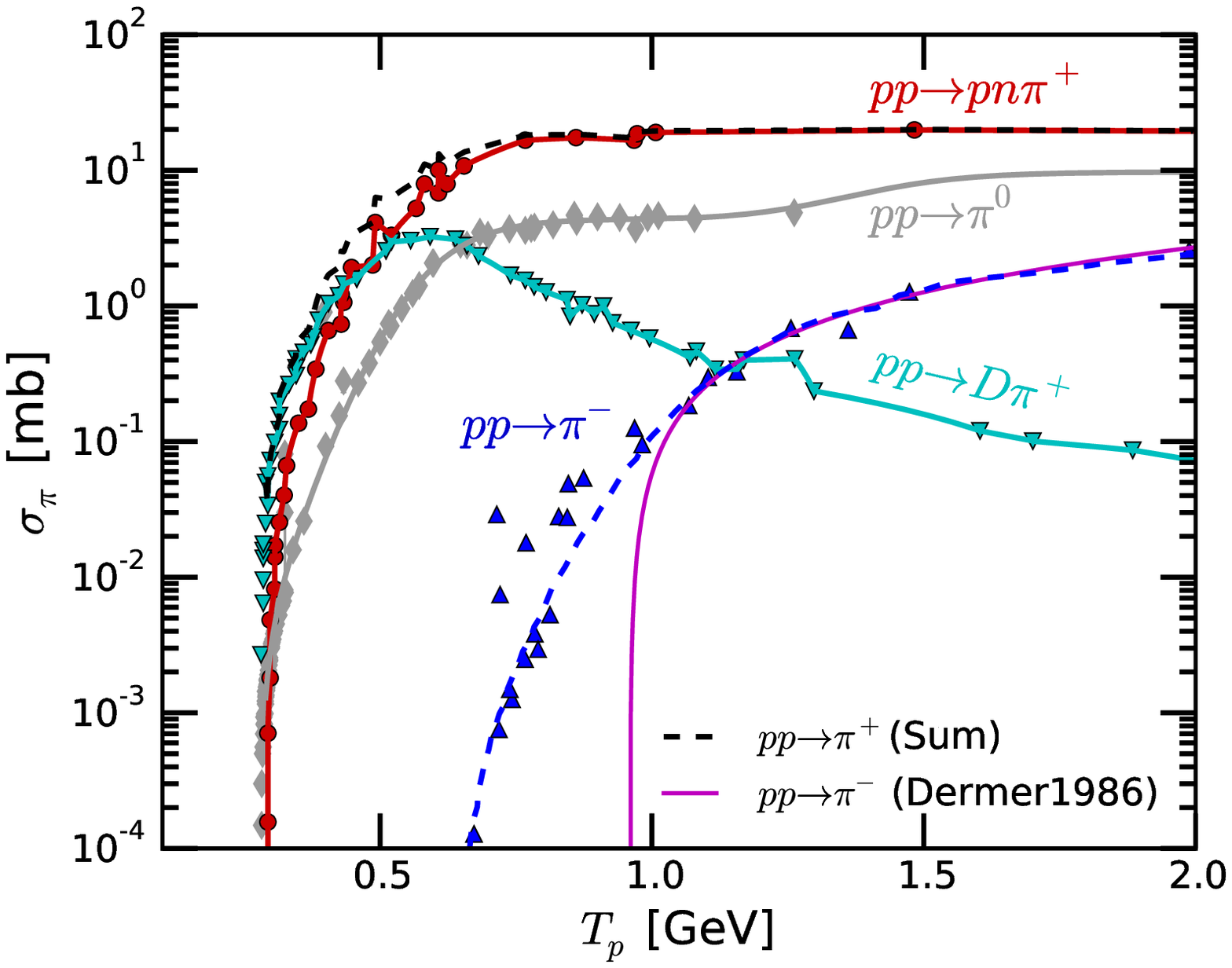}
\caption{$\pi$-meson production cross-sections  as a function of incident proton kinetic energy  $T_p<2$~GeV.  The most important channels for $\pi^+$-meson production are the $pp\to pn\pi^+$ (red line) and the $pp\to D\pi^+$ (cyan line) channels. The experimental data points are from  \cite{Machner1999}; the black dash line shows the total $pp\to\pi^+$ cross section. The cross-sections for the $pp\to\pi^0$ and $pp\to\pi^-$  reactions are shown with gray and blue colours, respectively.  The parametrisation for the $pp\to\pi^0$ and the respective compiled data are taken from \citet{kafexhiu14}.  The experimental data for $pp\to\pi^-$ and the eye-guiding blue dash line are taken from \citet{Skorodko2009}. For comparison, the \cite{dermer1986} parametrisation  for $pp\to\pi^-$  cross-section  is also shown. This parametrisation describes well the high energy data, however,  below 
$T_p<1$~GeV  \label{fig:XSPions} it deviates significantly from measurements.}
\end{figure}

\begin{figure}
\centering
\includegraphics[scale=0.54]{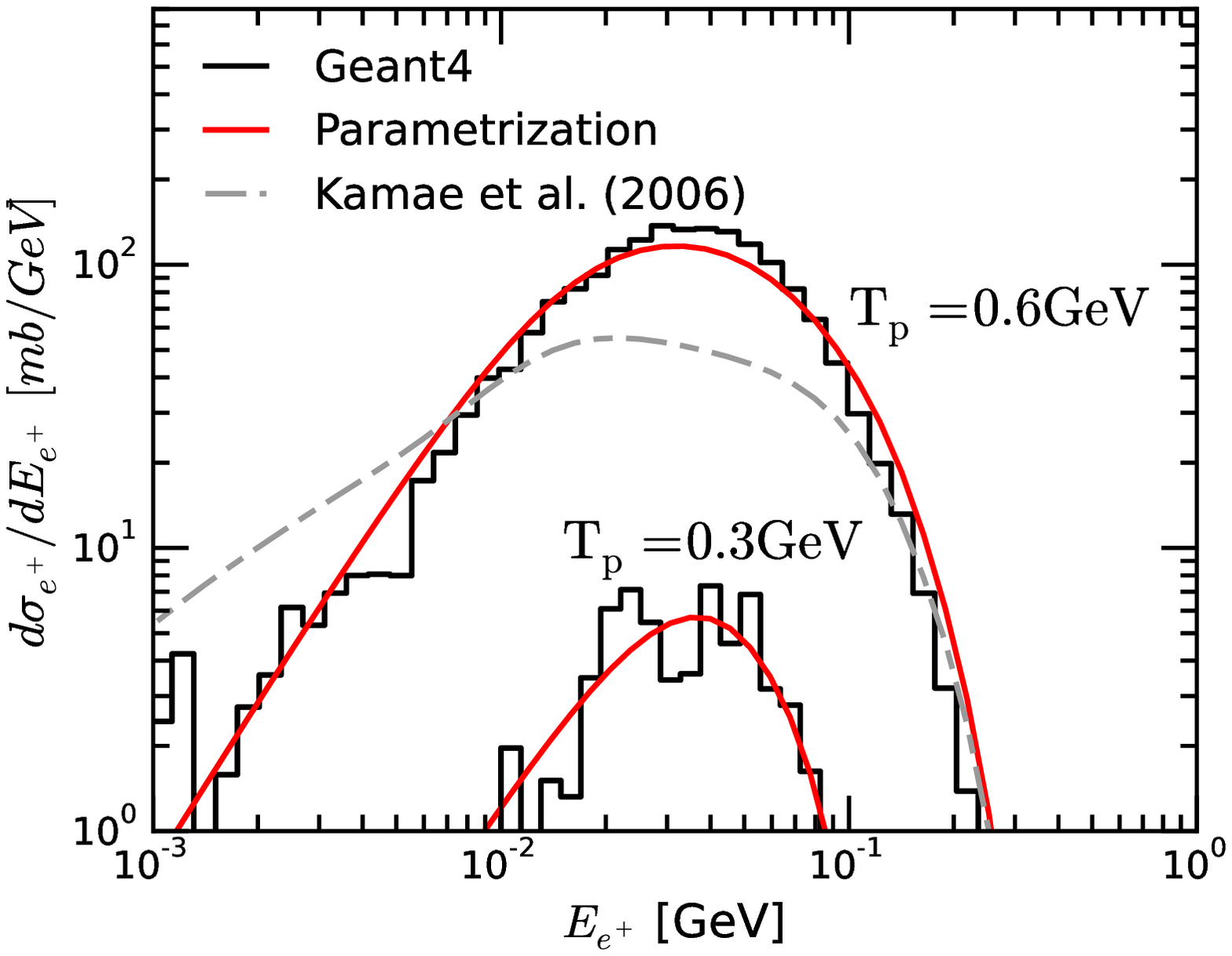}
\caption{The energy distribution of secondary positrons for  incident proton energies
$T_{\rm p}\approx0.3$ and 0.6~GeV. The histograms are obtained from simulations using the Geant4  tools, while the red lines correspond to   parametrisations  for 
$\pi$-meson production given by  Eq.~(\ref{eq:fx}) of Appendix~\ref{ap:Pip}. The gray long dashed  line is calculated using  the parametrization of  \cite{Kamae2006} for $T_{\rm p}\approx0.6$~GeV.  \label{fig:SpecPions}}
\end{figure}

Although, the process $pp\to\pi^0\to2\gamma$ is the dominant  \gray production channel, the secondary electrons can significantly contribute to the overall $\gamma$-radiation, especially below 100 MeV.  
%
These secondary electrons are included in the parametrisations  of  \cite{kelner06} and 
\cite{Kamae2006}, however,  they do not provide adequate accuracies at  low energies,  $T_p<0.5$~GeV. Meanwhile, this is the most important energy interval for contribution of the bremsstrahlung of secondary electrons to the $\gamma$-ray spectrum  below the $\pi^0$-decay bump. 
Below  we present our 
calculations of $e^\pm$ production based on the cross sections from the Geant4 toolkit \citep{Geant42003,Geant42006}. 
We adopt here the \textit{FTFP\_BERT} hadronic interaction model that implements the Bertini-style cascade model at low and intermediate energies, $T_p \leq5$~GeV, 
and the FRITIOF string model at higher energies. To make the computations of the secondary $e^\pm$ spectra rather convenient, we parametrise the $\pi$-meson production 
differential cross section in the following form:
\begin{equation}\label{eq:fx}
\frac{d\sigma_\pi}{dx} = \sigma_\pi\times f(x),
\end{equation}
where $x=T_\pi/T_\pi^{\rm max}$;  $T_\pi$ and $T_\pi^{\rm max}$ are the $\pi$-meson kinetic energy and its maximum kinetic energy (in the laboratory frame), respectively; $\sigma_\pi$ is the total cross-section of  the corresponding channel of pion production. The function $f(x)$ is the normalised pion energy distribution, i.e.  the integral of $f(x)$ from 0 to 1 is set to be one. In Appendix~\ref{ap:Pip} we offer a convenient  presentation of this function.   

The experiment data of  integrated cross sections of different channels 
of pion production at kinetic energies of protons  $T_p\leq2$~GeV 
are shown in  Fig.~\ref{fig:XSPions}. For higher energies, we use the
parametrisation  based on the approximation of the 
pion average multiplicity using  experimental data from
 \cite{Golokhvastov2001}. For this parametrisation, the pion average yield is expressed 
as $\left\langle n_\pi \right\rangle = 0.78\,F-1/2 + \varepsilon$, where $F=(w-2)^{3/4}\,w^{-1/4}$ and  $w=\sqrt{s}/m_p$.  Here $s$ is the total center of mass energy squared and $m_p$ is the proton mass. The cross sections are  calculated as  $\sigma_\pi=\sigma_{pp}\left\langle n_\pi \right\rangle$ where $\sigma_{pp}$ is the inelastic $pp$ cross section given in \cite{kafexhiu14};
$\varepsilon=$ 0, 1/3 or 2/3 for $\pi^-$, $\pi^0$ and $\pi^+$, respectively. 

The experimental cross-sections of  $\pi$-meson production   at low energies taken from 
\cite{Machner1999, Skorodko2009,kafexhiu14}, are shown in Fig.1,  together with 
the parameterisations presented in Appendix~\ref{ap:Pip}. It is seen that
the $\pi^-$ production cross-section below 2~GeV is very small compared to the 
production of  positive and neutral pions. The reason is that the negative pions are 
not produced through the 
$pp\to\Delta(1232)$-resonance. 

\begin{figure}
\centering
\includegraphics[scale=0.63]{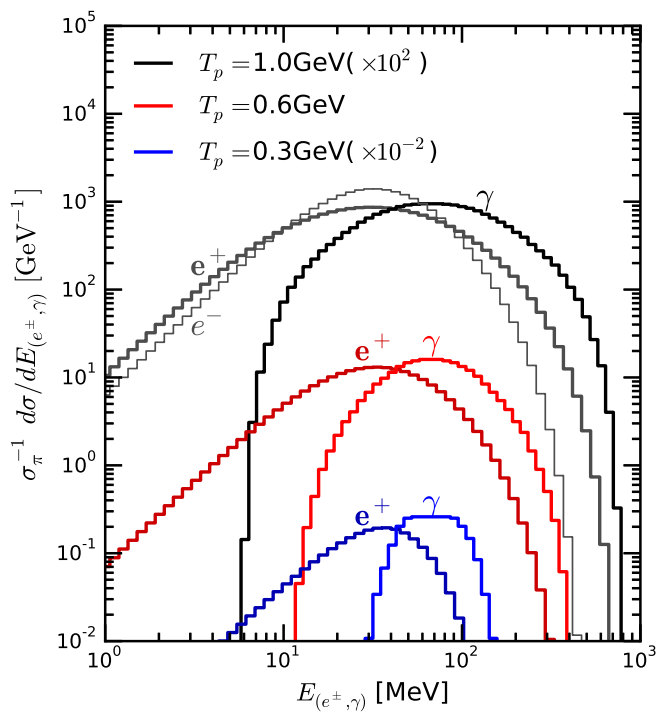}
\caption{The normalized differential energy distributions of  electrons, positrons, and
$\gamma$-rays as products of decays  of $\pi$-mesons  for three different energies of the incident protons  $T_{\rm p} =$~0.3, 0.6, and 1~GeV computed using Geant4.  
At low energies, the electron production is severely suppressed (the corresponding curves do not appear on the plot), while at $T_{\rm p}=1 \ \rm GeV$ the contributions of electrons and positrons become comparable.  The label on the top of each histogram shows the relevant channels. \label{fig:Specepg}}
\end{figure}

The positrons and electrons are produced at decays of the secondary 
$\pi$-mesons. The kinematics of such reactions are 
described in \citet{scanlon65, dermer86} and \citet{kelner06}.
Fig.\ref{fig:SpecPions} shows the positron energy distributions calculated for 
$T_{\rm p}=0.3$~GeV   and  0.6~GeV. The 
histograms represent the results of Geant4 Monte Carlo simulations,  while the solid curves represent the parametrisations  of this work.  For comparison, the parametrisation of  \cite{Kamae2006}  for  
$T_{\rm p}=0.6$~GeV is shown as well.  One can see that at energies 
of positrons less than  0.1~GeV, the parametrisation 
of  \cite{Kamae2006} significantly deviates from  the Geant4 simulations. 
Fig.~\ref{fig:Specepg}  shows the secondary $e^\pm$ and \gray production spectra 
at different kinetic energies  of the incident proton obtained from  Geant4 Monte Carlo simulations.

\begin{figure}
\centering
\includegraphics[scale=0.5]{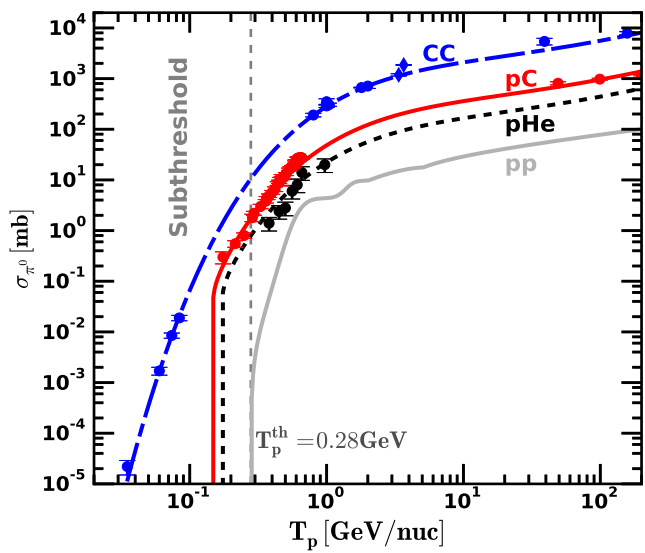}
\caption{$\pi^0$ production cross sections for the  $p+p$, $p+{\rm^4He}$, $p+{\rm^{12}C}$ and 
${\rm^{12}C}+{\rm^{12}C}$ interactions. The experimental points are shown together with  the paramerizations of cross-sections from \cite{kafexhiu2016}. }
 \label{fig:ABSpecPions}
\end{figure}
\subsection{contribution from heavy nuclei}
The nuclei heavier than hydrogen can significantly contribute to the formation of the $\gamma$-ray spectrum. 
The production of secondary particle  at  high energy interactions,
can be described  in terms of Glauber's multiple scattering theory \citep{Glauber1955,Franco1966,Glauber1970}. In this  model, the nuclear interactions are considered as a sequence of binary nucleon--nucleon collisions. Correspondingly,  the energy distributions of  secondary particles  are  approximated  by functions relevant to  the nucleon-nucleon interactions. Therefore,  for the treatment of $\gamma$-rays produced 
at interactions of  Galactic CR protons and nuclei  with the interstellar gas,  
typically the emissivity of $\gamma$-rays from $pp$ interactions  
is multiplied by the so-called nuclear enhancement factor $\kappa \approx 1.5-1.8$ \citep{Cavallo1971,Stephens1981, dermer1986, Mori2009,kafexhiu14}.

At lower (subrelativistic) energies,  the  nuclear interactions no longer can be treated as a sequence of binary nucleon--nucleon interactions.  In particular, the collective effects,  among other things, allow the nucleons inside the nucleus to produce secondary particles at energies below the 
kinematic thresholds of  the pion production  \citep[see e.g.][]{Cassing1990,Metag1993}.  Despite the lack of  
a self-consistent theory of the 
\textit{subthreshold pion} production, the cross-sections of many heavy ion reactions 
are well studied experimentally
\citep{Cassing1990,Metag1993}. Recently, these cross-sections have been 
parametrised by simple analytical expressions  \citep{kafexhiu2016}.  In Fig.\ref{fig:ABSpecPions} the available experimental data are shown together with parametrisations of  the neutral pion production cross-sections for  p -$^4$He, p -$^{12}$C, and $^{12}$C - $^{12}$C interactions. In addition to  $\gamma$-rays 
from the decays of  "subthreshold"  $\pi^0$-mesons,  at low energies,
 typically between 30 and 100 MeV,   a $\gamma$-ray continuum 
 is formed  through  the so-called direct \textit{hard photon} channel \citep[see e.g.][]{Cassing1990,Metag1993}.  For the typical CR spectra and the cosmic abundances, the  contribution of these two mechanisms below 100 MeV does not 
 exceed 10 percent  (when compared to the extension of the spectrum of 
$\gamma$-rays  from the  "nominal"  $\pi^0$ production). However in the case of "heavy" 
compositions of both CRs and the ambient gas  dominated by nuclei, these processes could play a non-negligible role in  the  formation of the overall  $\gamma$-ray spectrum  below 100~MeV.  The Cross sections for hard photon production are parametrised by \citet{kafexhiu2016}. 

\begin{figure*}
\centering
\includegraphics[scale=0.45]{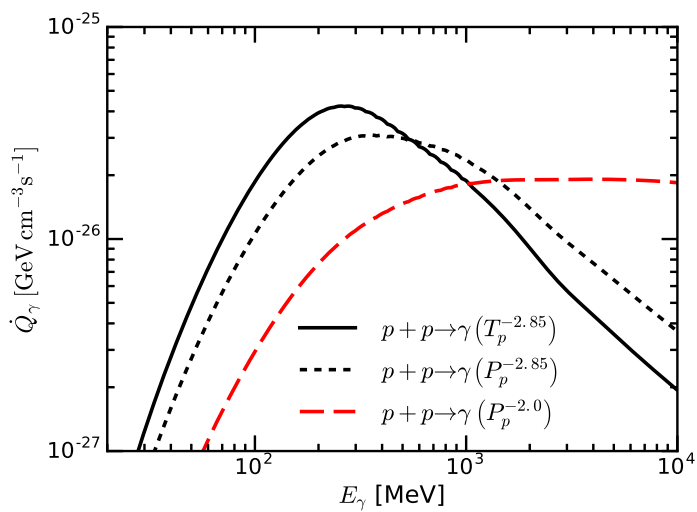}
\caption{Emissivities of $\gamma$-rays from decays of $\pi^0$-mesons produced
in $pp$ interactions. Different line styles  are calculated for three distributions of accelerated 
protons:  (1)red dashed:  power-law  in momentum with $\alpha_{\rm p}=2.0$,   
(2) black dased:  power-law  in momentum with $\alpha_{\rm p}=2.85$, (3) black solid:  power-law   in kinetic energy with $\alpha_{\rm p}=2.85$, respectively.   The proton spectra are normalised in a way that 
the energy density $w_{\rm p}(\geq 1 \  \rm GeV)=1 \  \rm eV/cm^3$.  The density of the hydrogen gas $n_{\rm H}=1 \  \rm cm^{-3}$.}
\label{fig:pp}
\end{figure*}

In astrophysical environments,  the accelerated particles are generally 
described by a power-law distribution either in kinetic energy 
or in momentum.  Apparently, at relativistic energies  these two presentations 
essentially coincide.  Note that at these energies  
the spectrum of $\gamma$-rays from the decay of $\pi^0$-mesons almost mimics the spectrum of parent particles. In particular, the $\gamma$-ray spectrum also  behaves as  power-law but, because of the slight increase of the $\pi^0$ production cross-section,  it appears to be a bit harder.  Namely, over the energy interval of a few decades,  $\alpha_\gamma=\alpha_{\rm p} +\Delta \alpha $,  with $\Delta \alpha  \approx  0.1$.  On the other hand, the 
$\gamma$-ray spectrum around  the $\pi^0$-bump strongly depends on the spectrum of parent particles at low energies.  It can be seen in  in Fig.~\ref{fig:pp} where the $\gamma$-ray emissivities  are shown.  The curves in Fig.~\ref{fig:pp} correspond to three spectra of accelerated 
protons:  power-law in momentum with $\alpha_{\rm p}=2.0$  and 2.85, and in kinetic energy with 
$\alpha_{\rm p}=2.85$.  The sharpest  distribution in the bump region 
appears in the case of the proton spectrum $\propto  T_p^{-2.85}$.  This is naturally explained by the 
excess of nonrelativistic protons below 1~GeV. 
On the other hand,  in the case of the hard proton distribution like  $p^{-2}$  (or harder),  
the bump in  the $\gamma$-ray SED  disappears at all. 

\begin{figure*}
\centering
\includegraphics[scale=0.45]{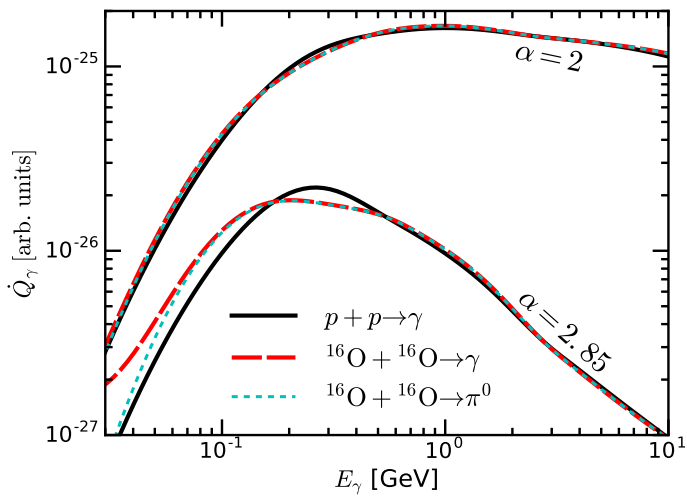}
\caption{Emissivities of $\gamma$-rays from decays of $\pi^0$-mesons produced
in $pp$ (solid lines) and $\rm ^{16}O ^{16}O$ (cyan dashed lines) interactions for two energy distributions of particles:  power-law in kinetic energy, $T^{-\alpha}$ with $\alpha=2$ and 
$\alpha=2.85$, (for both $p$ and $\rm ^{16}O$ projectiles).  Also plotted is the total  $\gamma$-rays produced in $\rm ^{16}O ^{16}O$ (red dashed lines) interactions, including the hard photon channel.  The emissivities of $\gamma$-rays are shown in arbitrary units 
and are normalised at 1 GeV to demonstrate the differences  between the spectra of 
$\gamma$-rays from two reactions.
}
\label{fig:PandO}
\end{figure*}

\begin{figure*}
\centering
\includegraphics[scale=0.45]{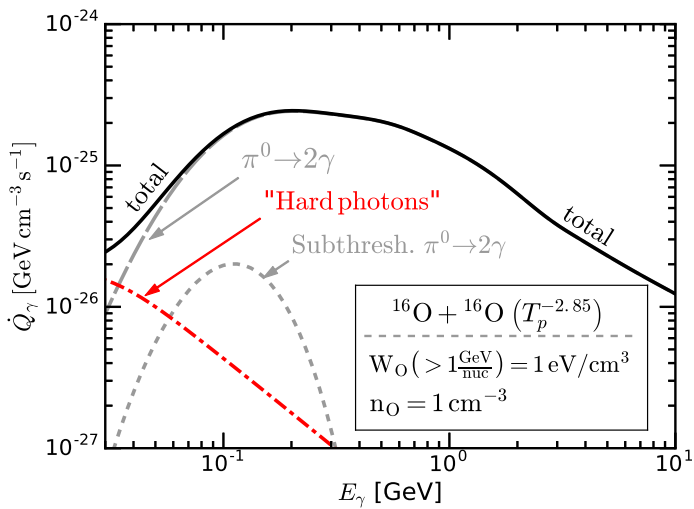}
\caption{Emissivities of $\gamma$-rays  produced
in $\rm ^{16}O ^{16}O$ interactions . The three curves correspond to three radiation channels: $\gamma$-rays from (1) gray long dashed line: decays of  ``nominal''  $\pi^0$-mesons, 
(2) red dash-dotted line: "hard photons", and (3)  gray short dashed line: decays of "subthreshold"  $\pi$-mesons. The spectrum of nonthermal nuclei $\rm ^{16}$ is assumed a power low in kinetic energy with  $\alpha_{\rm p}=2.85$. 
For the energy spectrum of CRs and the density of gas  the following normalisations are used: $w_{\rm p}(\geq 1 \  \rm GeV/nuc)=1 \  \rm eV/cm^3$;  $n_{\rm O}=1 \  \rm cm^{-3}$.
}
\label{fig:OO}
\end{figure*}

At heavy  nuclear interactions, the $\pi^0$-decay bump in the $\gamma$-ray spectrum 
is similar  but not identical to the  analogous feature  in  the spectrum of 
$\gamma$-rays from $pp$ interactions.  It is seen in  Fig.\ref{fig:PandO} where the spectrum of  
$\gamma$-rays produced in $pp$ interactions is  compared with  the spectrum of $\gamma$-rays  produced  at  interactions  when both the projectile and target particles are  
the same $\rm ^{16}O$ nucleus.  The difference is  not dramatic ( within 20 percent ) above 100 MeV, although the differential cross-sections of production of $\pi^0$-mesons in $pp$ and $\rm ^{16}O^{16}O$ interactions  deviates significantly (up to by factor of three) at lowest energies.   This is because the integrated $\gamma$-ray spectrum from the decay process of $\pi^0$-mesons  smears out most of the difference in the spectrum of the $\pi^0$-mesons. 
At low energies, the  additional contributions  from the "hard photon" continuum and "sub threshold  $\pi^0$-meson" production channels can be significant, especially in the case of 
a steep 
energy distribution of projectiles extending down to  100 MeV/nuc. In 
Fig.\ref{fig:OO}, the calculations  are performed for the 
power-law spectrum in kinetic energy of $\rm ^{16}O$ (the power-law index 2.85)  using the following normalisations: the kinetic energy density of  the 
$\rm ^{16}O$ nuclei above 1 GeV/nuc, $w_{\rm O} =1 \ \rm eV/cm^3$,  and the number
density of the target $\rm ^{16}O$ nuclei $n_{\rm O} = 1 \ \rm cm^{-3}$.  
Generally, the chemical composition of the diffuse gas is dominated  
by atomic or molecular hydrogen. However,  in some environments 
the heavy nuclei can dominate over the hydrogen.  For example, this is the case of
the young supernova remnant Cas A   where the gas consists of heavy elements, 
especially of oxygen \cite{cas_composition}.  It seems natural to expect that CRs 
accelerated in the shell of this supernova remnant,  will be dominated by $\rm ^{16}O$ as well,  especially in the reverse shock region \citep[see, e.g., ][]{zirak14}. 
Therefore, the calculations shown in   Fig~\ref{fig:OO}  present not only an academic interest but could be realised in certain astrophysical scenarios.  

\begin{figure*}
\centering
\includegraphics[scale=0.55]{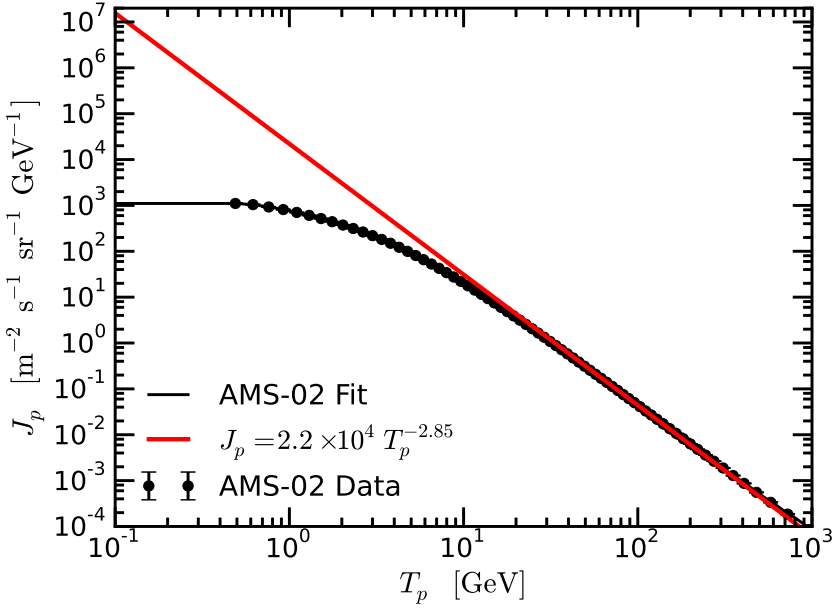}
\caption{The Cosmic Ray proton spectrum reported by AMS \citep{AMS2} (experimental points) 
shown together with  two extrapolations (see the text).}
\label{fig:AMS}
\end{figure*}

\begin{figure*}
\centering
\includegraphics[scale=0.33]{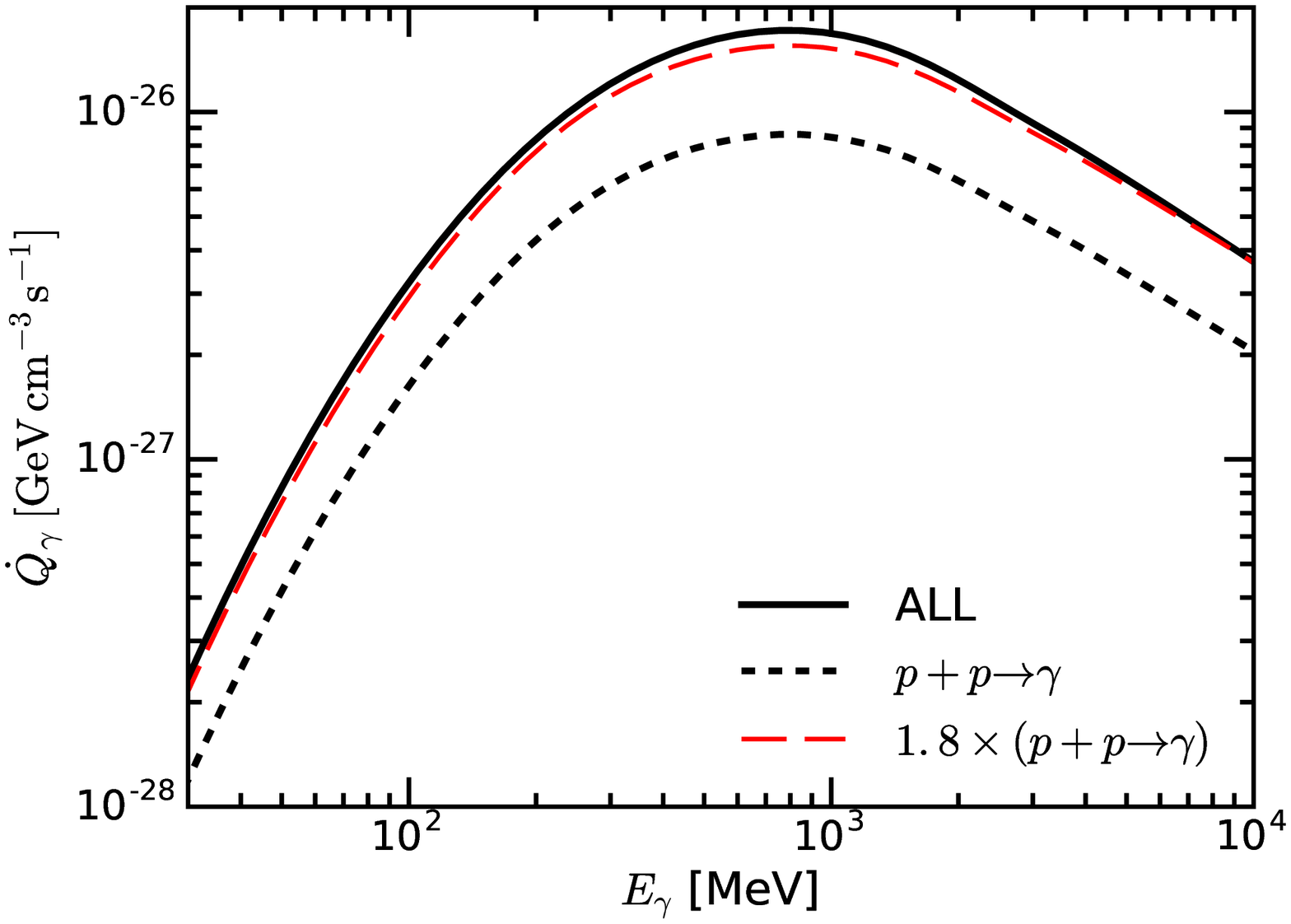}
\includegraphics[scale=0.33]{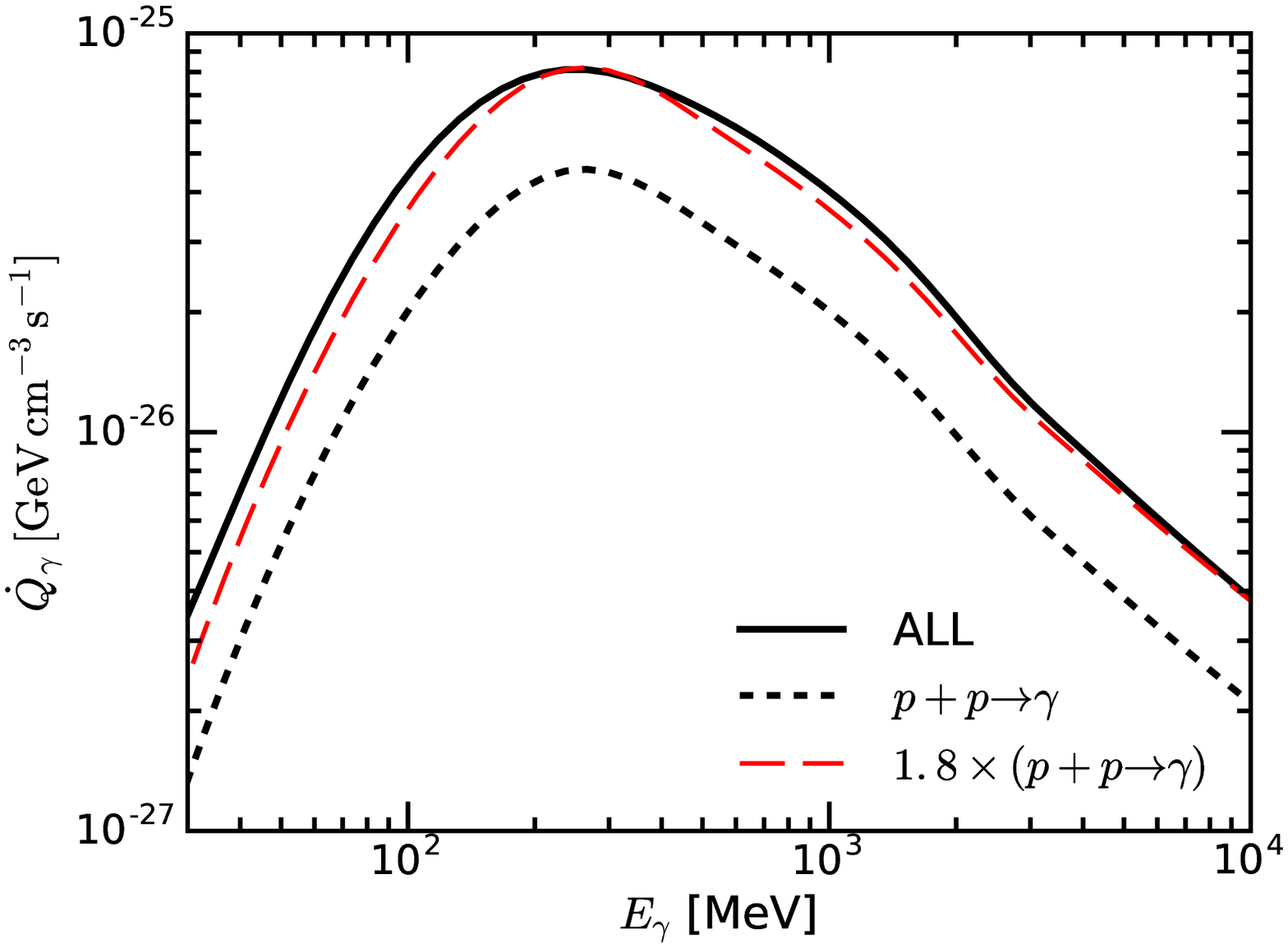}
\caption{Emissivities of $\gamma$-rays from interactions of Galactic CRs with the interstellar gas. 
{\ left panel}: the overall $\gamma$-ray emissivity is  shown by the solid line, while the emissivity due to $pp$ interactions is shown by the dashed line. The red line corresponds to the $pp$ emissivity  enhanced by the factor of $\kappa=1.8$ in order to match the overall emissivity at energies above 3~GeV.  For composition and the  CR proton flux are used the AMS measurements. 
{\it right panel}: the same as the left panel, except for the proton spectrum; it  is assumed the 
same  AMS  proton spectrum at high energies but extrapolated to low energies as a single power-law with the same power-law index 2.85. }
\label{fig:GCR}
\end{figure*}

The contributions of the "hard photon" continuum  and the "sub-threshold" $\pi^0$-decay 
$\gamma$-rays is  dramatically reduced when protons dominated 
over nuclei both in the projectile and target particles. This is the case of interactions of Galactic CRs with the interstellar gas.  For the  energy spectrum and the chemical compositions of CRs  
below we use the recent  measurements by the AMS detector \citet{AMS1,AMS2}.  The energy spectrum of CRs  reported by the AMS collaboration \citep{AMS2} is shown in Fig.\ref{fig:AMS}. 
The proton spectrum  at high energies, $E \geq 10$~GeV  is described as a power-law with 
$\alpha_{\rm p}=2.85$.
But  at low energies, below 10 GeV it becomes flatter. This  explains the rather shallow distribution of $\gamma$-rays in the $\pi^0$-bump region; see Fig.\ref{fig:GCR}a. 
However, since the  suppression of the proton flux at low energies could be a result of a local effect, e.g. due to the propagation effects in the Solar System, in  Fig.\ref{fig:GCR}b we show 
the $\gamma$-ray luminosities for the proton spectrum which we extrapolate, starting at the energy of 10 GeV  to subrelativistic  energies as a single power-law in kinetic 
energy with the same power-law index $\alpha_{\rm p}=2.85$ (see Fig.\ref{fig:AMS}).  
The solid curves in  Figs.\ref{fig:GCR}a,b represent  the overall gamma-ray emissivities 
contributed by interactions with an involvement of all nuclei from  both the interstellar gas and CRs. For comparison,  we show also the emissivities of $\gamma$-rays produced only in
$pp$ interactions (dashed lines).  One can see that the contribution of all nuclei from the Galactic CRs and the interstellar gas is significant; it almost doubles the contribution of protons. 
In  Figs.\ref{fig:GCR}a,b   we  introduce  curves  obtained by multiplying the contributions 
of protons  to the $\gamma$-ray emissivity by a factor  of $\kappa=1.8$ to enhance it to the level 
of the overall $\gamma$-ray luminosity above 3 GeV.  One can see that the  spectral shapes of  the "protonic" and overall $\gamma$-ray emissivities are quite similar.  The fluxes at 
any  point  from 100~MeV to 10~GeV  does not  exceed  10 \%.

Finally, we should mention that at low energies,  the nuclear  interactions, with involvement of nuclei of both CRs and the ambient
gas,  result in intensive  $\gamma$-ray line production \citep{Murphy2009}.  The   prompt 
de-excitation $\gamma$-ray lines  contain unique information 
CRs at energies  less that 100 MeV/nuc.  The contribution of this channel to 
the $\gamma$-ray production is limited by  energies below 
10 MeV, therefore they do not appear in   and Figs.\ref{fig:OO}  and \ref{fig:GCR}.
Even for very "heavy" composition and very soft energy distributions of 
CRs,  the impact of de-excitation lines on the $\gamma$-ray spectrum in the region of the $\pi^0$-decay bump, is negligible.

\section{\gray emissivities}

\begin{figure*}
\centering
\includegraphics[scale=0.65]{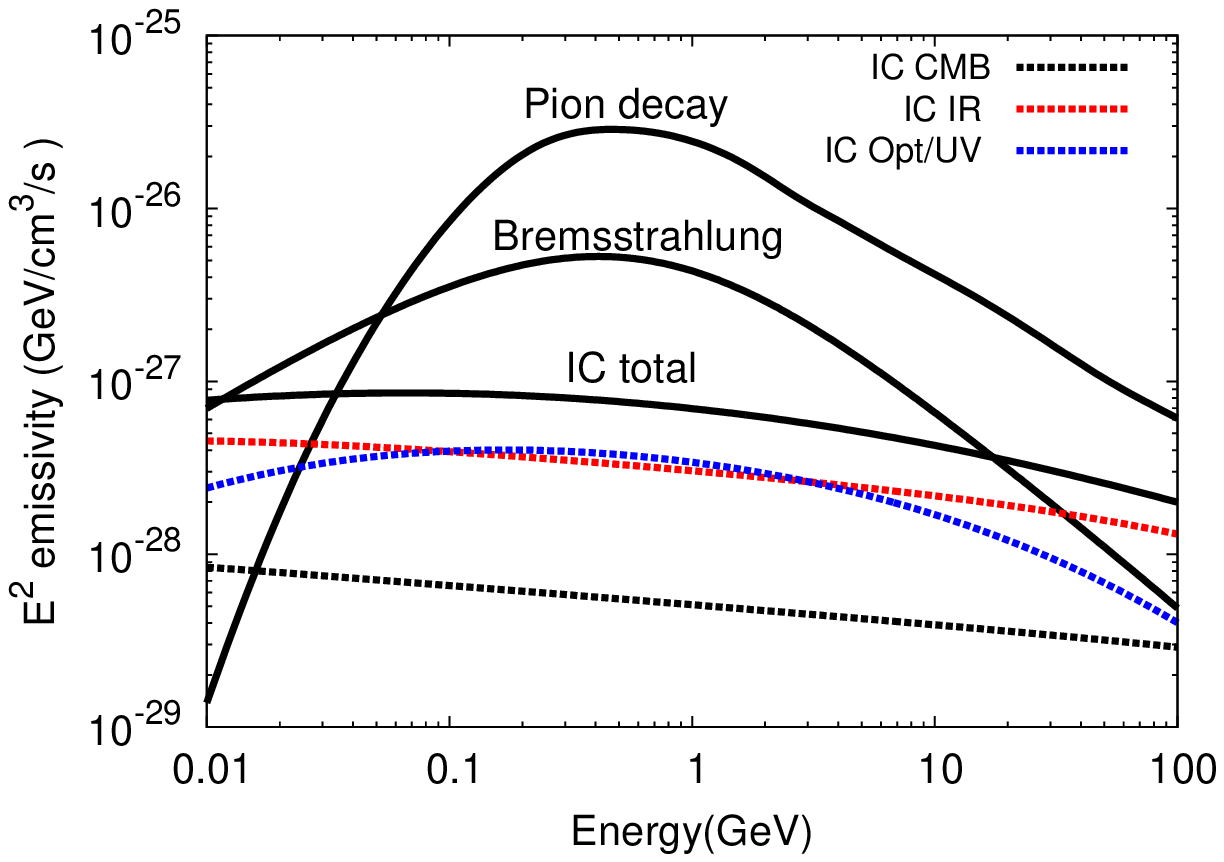}
\caption{ The \gray emissivities in the interstellar medium.  The energy spectra of CR electrons and protons are taken from \citet{casandjian15}.  The ambient gas number density  is  
$1~\rm cm^{-3}$.  The background radiation fields consist of three components: (1) CMB with an  energy density of $0.24 ~\rm eV/cm^3 $ and temperature of $2.7~\rm K$, 
the optical/UV component which is modelled as a grey body radiation  with an  energy density of $2 ~\rm eV/cm^3 $ and temperature of $5000~\rm K$  and (iii)  IR component which is approximated as a grey body component with an  energy density of $1 ~\rm eV/cm^3 $ and temperature of $100~\rm K$ . } 
\label{fig:ism}
\end{figure*}

Besides the hadronic processes CRs  can also produce high energy \grays via   (i) electron bremsstrahlung, and (ii)  inverse Compton (IC)  scattering  of electrons.  
We illustrate the contributions of  these channels for the interstellar medium (ISM) in Fig.\ref{fig:ism}, where we assume an ambient density of $1~\rm cm^{-3}$ and a background photon field as described in the previous section. The proton and electron spectra used in calculations are taken from \citet{casandjian15}, where the local interstellar spectra (LIS) of CRs have been derived from local \ion{H}{i} emissivities. The results show that from $100~\rm MeV $ up to $100~\rm GeV$ the pion decay dominates in the \gray production. At lower energies the bremsstrahlung becomes more important.  In the LIS there are  low energy cutoffs in both electron and proton spectrum, which produce  corresponding features in the \gray spectra.

\begin{figure*}
\centering
\includegraphics[scale=0.65]{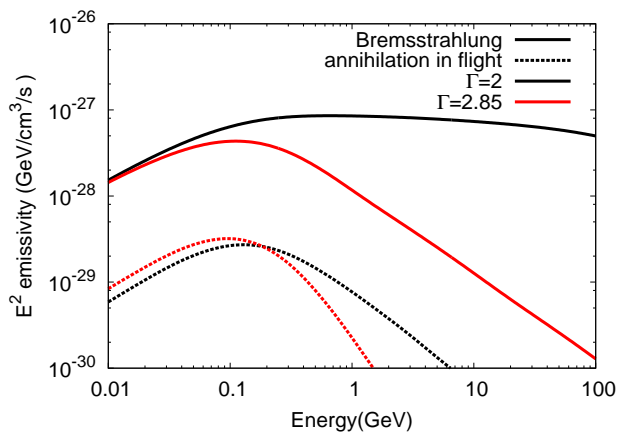}
\caption{ The \gray  emissivities  of secondary electrons and positrons through  Bremsstrahlung and annihilation in flight. The electrons spectra  are calculated  in  the saturation regime for  
two power law distributions of protons in momentum  with indices 2.0 and 2.85 normalised 
to $1 ~\rm eV ~\rm cm^{-3}$ above 1 GeV. The  gas number density is 
$n=1 \ \rm cm^{-3}$. } 
\label{fig:ani}
\end{figure*}

While at higher density regions both the pion decay and the bremsstrahlung \gray
 emissivities increase with the ambient gas density, the ICs contribution remains unchanged if the radiation fields are kept constant.  Thus,  in such environment we can ignore the IC scattering of electrons almost 
in the entire $\gamma$-ray band.  On the other hand,  the  relativistic positrons can 
contribute to the $\gamma$-ray emission not only through the bremsstrahlung, but also 
through the process called annihilation in flight \citep{aif}.  In particular, 
at  the presence of a primary component of  relativistic positrons 
(e.g. from pulsars),  the process of annihilation in flight can significantly contribute to the overall 
diffuse galactic $\gamma$-ray background below 100 MeV  \citep{AAdif}.  
In this regard, one may expect some contribution also from the secondary positions, especially given the high $e^+/e^-$ ratio from $\pi^\pm$-decays at low energies.  However, because of the very hard spectra of positrons produced in $pp$ interactions below 100~MeV, the contribution of this channel appears quite modest. This can be seen in  
Fig.\ref{fig:ani} where  the contributions of bremsstrahlung and the annihilation in flight from secondary electrons and positrons are plotted.  For both steep and hard spectra of primary protons, the contribution of the annihilation of positions in flight  is an order of magnitude below the contribution of bremsstrahlung.

 Below we compare the  emissivities of $\gamma$-rays originating directly from the interactions of 
 primary CRs with gas and  from the secondary ($\pi^\pm$-decay) electrons and positrons.  
 We consider  an environment typical for the interstellar medium.  Since we are interested in the energy around the pion-decay bump, the contribution of the IC scattering of electrons is not significant. Therefore, for  simplicity, we will ignore the IC radiation channel  For higher density environments, e.g. in dense molecular clouds,  the  IC contribution can be safely ignored up to  1~TeV.

\begin{figure*}
\centering
\subfigure[][$\Gamma=-2$ in momentum]{
\includegraphics[scale=0.65]{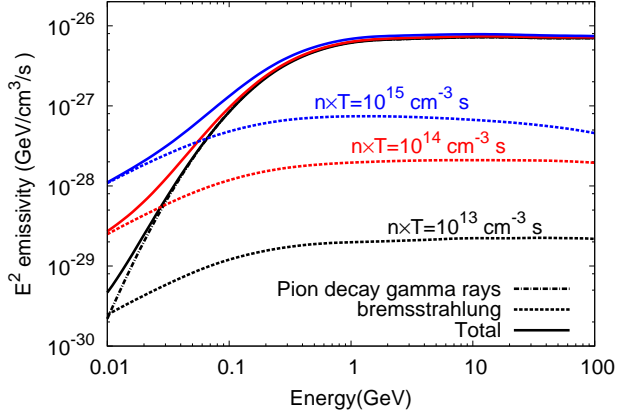}
}
\subfigure[][$\Gamma=-2.85$  in momentum]{
\includegraphics[scale=0.65]{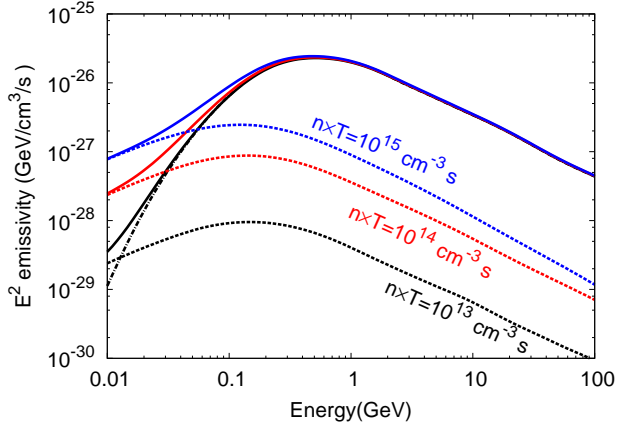}
\label{OrionB:4}
}
\caption{The \gray emissivity of the secondary electrons and positrons and parent protons  at different epochs 
characterised by the parameter $n\times T$  (in unit $cm^{-3} s$):  $10^{13}$ (black curve ), $10^{14}$ (red curve),  $10^{15}$ (blue curve). 
different $n\times T$ value. The distribution of the parent protons is assumed saturated with an injection power-law spectrum in momentum with indices $\Gamma=2$ (left panel) and $\Gamma=2.85$ (right panel). The proton density are normalized such that the energy density above 1~GeV is $1~\rm eV/cm^3$. The  gas number density is 
$n=1 \ \rm cm^{-3}$.}
\label{fig:sec}
\end{figure*}

In Fig.\ref{fig:sec} we show the $\gamma$-ray emissivities contributed directly by CRs though the production and decay of $\pi^0$-mesons, and through the bremsstrahlung of the secondary electrons and positrons. The curves in the figure correspond to different  epochs characterised by the parameter
$n\times T$:  $10^{13}$ (curve 1), $10^{14}$ (curve 2), 
  $10^{15}$ $\rm sec/cm^3$ (curve 3). One can see the contribution of secondary electrons increases with 
$n\times T$. But at some stage, when the parameter $n\times T $ exceeds 
$10^{15}~\rm s/cm^3 $, the emissivities achieves its maximum, i.e.  saturates.
While,  the contribution of secondary electrons to $\gamma$-rays above 
100 MeV is not significant,  at lower energies, e.g. at 30 MeV, the bremsstrahlung of secondary electrons can overcome the flux of $\pi^0$-decay $\gamma$-rays by an order of magnitude. 
Thus, in dense sources with effective confinement of low energy protons, the  effect of distortion of the $\pi^0$-decay $\gamma$-ray spectrum at energies of tens of MeV becomes significant.  This component can be revealed by future low-energy gamma-ray missions like e-ASTROGAM  \citep{astrogam}.


Finally,  we  investigated the contribution from  primary electrons  
by assuming  different e/p ratios. The calculations were performed for two 
extreme  regimes, namely, assuming uncooled and saturated distributions of particles. 
For the acceleration spectrum of both protons and electrons we assume 
power-law distributions in momentum with two spectral indices $\Gamma=$ 2 and 2.85. 
It is assumed that the power-law distributions of initial (= uncooled) particles continue down to 
kinetic energy 10 MeV.  As before  the gas number density is set $1 \ \rm cm^{-3}$, and the proton distribution is  normalised in a way that the energy density of protons is $1 \ \rm eV/cm^3$.   

\begin{figure*}
\centering
\subfigure[][$\Gamma=-2$ in momentum]{
\includegraphics[scale=0.65]{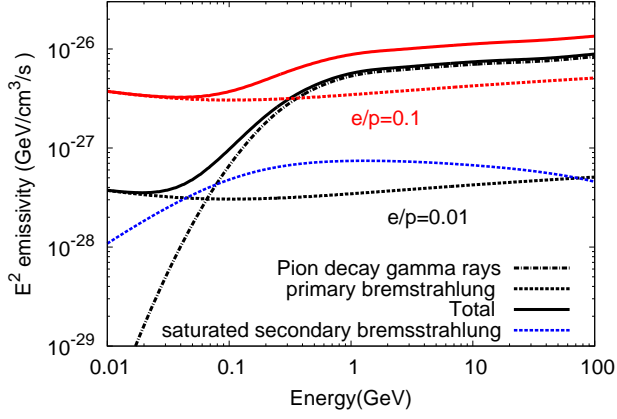}
}
\subfigure[][$\Gamma=-2.85$ in momentum]{
\includegraphics[scale=0.65]{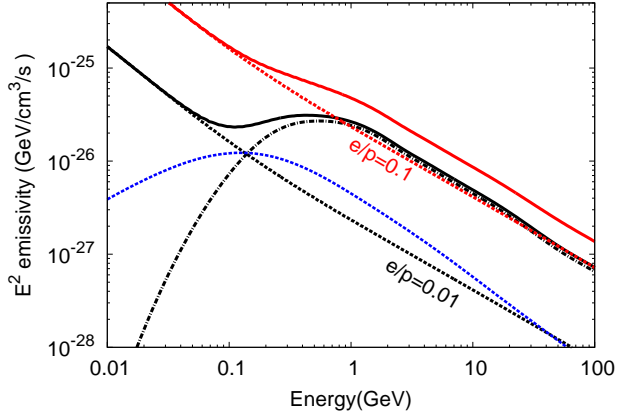}
\label{OrionB:4}
}
\caption{Gamma ray emissivities of  primary protons and electrons in the uncooled regime 
for two values of the e/p ratio: 0.1 and 0.01. 
It is assumed that both electrons and protons have  the same power-law distribution in momentum which is continues down to kinetic energy of particles of 10 MeV.  The  gas number density is 
$n=1 \ \rm cm^{-3}$.
Left panel: $\Gamma=2$; Right panel: $\Gamma=2.85$.  Also shown as blue curves are the contributions from saturated secondary electrons, which is the maximum possible contribution from secondaries. 
}
\label{fig:pri_uncool}
\end{figure*}

\begin{figure*}
\centering
\subfigure[][$\Gamma=-2$ in momentum]{
\includegraphics[scale=0.65]{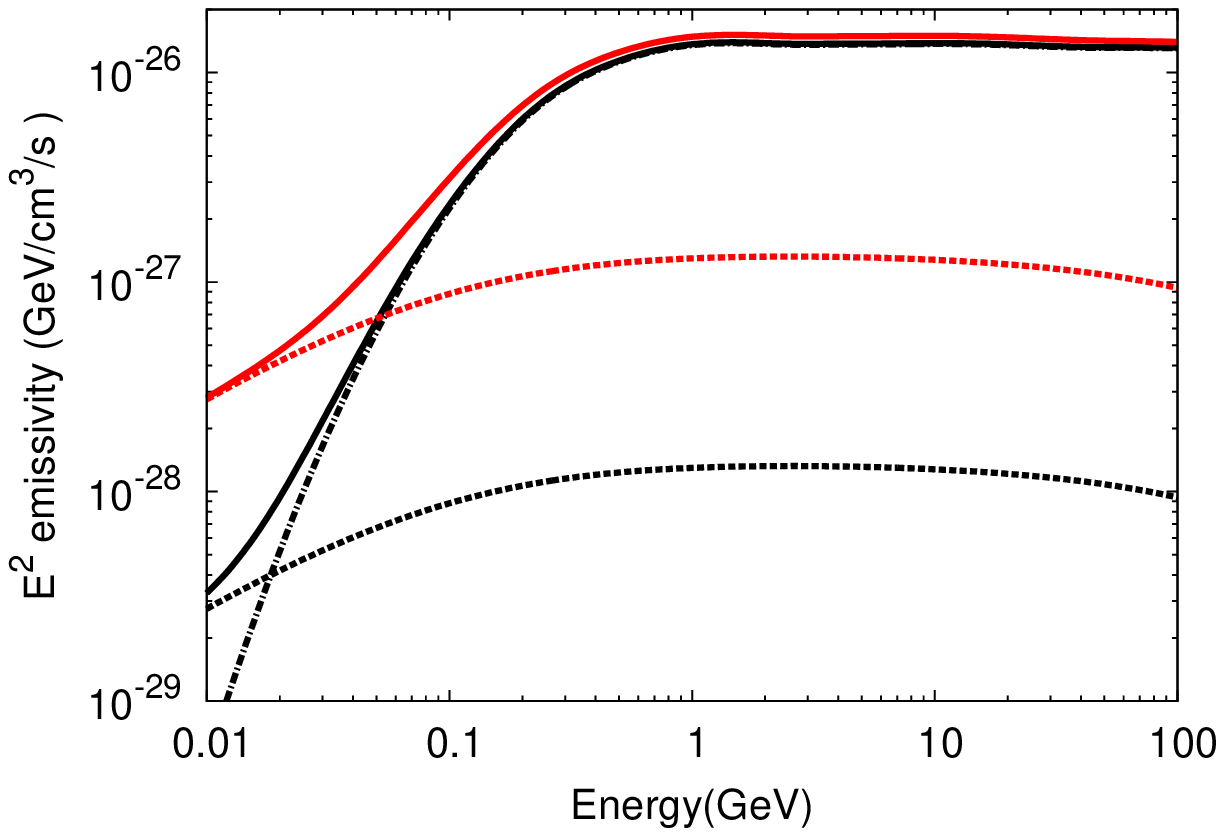}
}
\subfigure[][$\Gamma=-2.85$ in momentum]{
\includegraphics[scale=0.65]{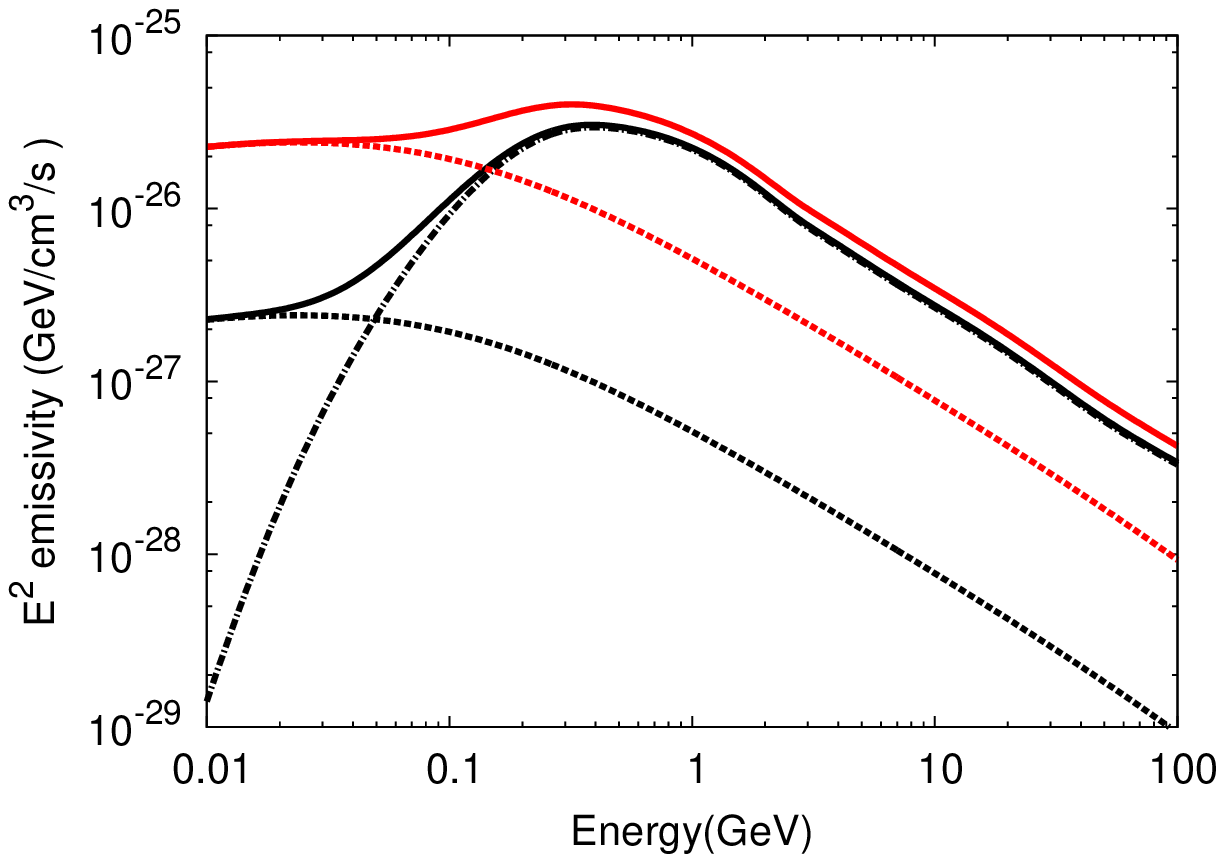}
\label{OrionB:4}
}
\caption{The same as in Fig.\ref{fig:pri_uncool}  but for the saturated regime} 
\label{fig:pri_sat}
\end{figure*}

The results calculated for two values of the ratio e/p=0.01 and 0.1, are shown in Fig.\ref{fig:pri_uncool}
and Fig.\ref{fig:pri_sat}.  In the case of uncooled particles, the bremsstrahlung 
\grays from primary electrons fill significantly the gap below the pion-decay bump. 
One can see that   even for the e/p ratio of 0.01  the contribution of primary electrons is 
pronounced, especially for the steep power-law distribution  with $\Gamma=2.85$.  
The effect is significantly less in the saturation regime. The contribution of  saturated primary electrons is
substantially  reduced which is explained by the suppression of these electrons due to the ionisation losses. The ionisation losses reduce also the contribution of the $\pi^0$-decay $\gamma$-rays, 
but for them the effect of ionisation losses  is  less significant.  {We also plot in  Fig.\ref{fig:pri_uncool}
and Fig.\ref{fig:pri_sat} the \gray emissivities from saturated secondary electrons and positrons, which is the maximum possible contribution from secondaries. We found that the \gray emissivities from secondary electron/positrons is comparable to that from primary electron/positrons in the saturated case even if the e/p ration is as high as 0.1. In the uncooled case, however, the \gray emissivities from primary electron/positrons dominate below 100 MeV when e/p is larger than 0.01.    }

\section{Summary}

The $\pi^0$-decay bump is a distinct spectral feature of $\gamma$-rays produced in interactions of 
CR protons and nuclei  with the ambient gas.  It is generally considered as a key signature for identification of the hadronic component of radiation, 
and, correspondingly,  for  probing the relatively low energy ($E \leq 10$~GeV/nuc) CRs in a broad range of astrophysical environments, from  stellar atmospheres and  supernova remnants to the interstellar medium and clusters of galaxies.  The shape of the $\pi^0$-decay bump is determined basically by the cross-section of  
$pp$ inelastic interactions and the energy distribution of nonthermal protons. The interactions of heavy ions 
can be responsible for a significant fraction of $\gamma$-ray production, but they do not change the spectral 
shape of the $\pi^0$-decay bump, except for the energy band below 100 MeV. 
For a heavy composition and soft spectral distribution of CRs  at  sub-relativistic energies,
the "hard photon" continuum   and the "sub-threshold"  pion production channels 
could significantly contribute to the overall $\gamma$-ray emission.  

At energies below  the $\pi^0$-bump,  bremsstrahlung of  secondary electrons and positrons, the products of charged $\pi$-mesons,  may provide 
a non-negligible  contribution to the radiation from hadronic interactions of CRs with matter. 
The significance of this indirect but unavoidable channel depends on the parameter  
$n\times T$, where $n$ is the number density of the ambient gas, and $T$ is the confinement time of CRs 
in the $\gamma$-ray production region. When the parameter 
$n\times T$ approaches  $5 \times 10^{15} \ \rm cm^{-3} s$,
the evolution of CR protons saturates, thus the source operates as a calorimeter. In this regime,  the contribution  of this channel below 100 MeV can be  as large as the contribution from the direct $\pi^0$-decay channel.  Therefore, the precise spectral measurements in the  energy interval between 10 MeV and 100 MeV may contain  an important information about the confinement of low energy cosmic rays, provided that the  spectrum of primary (directly accelerated) electrons is not significantly steeper than $E^{-2}$ and the ratio of primary electrons to protons, e/p,  does not exceed 0.01.



\begin{appendix}
\section{parametrisation of secondaries production cross section \label{ap:Pip}}
Geant4 simulations show that in $pp\to\pi$ process the  energy distributions for all three $\pi$-meson species appear to have a  similar shape,  thus we have parametrized them with a universal function as follows: 
\begin{equation}\label{eq:fx}
f(x) = \left\{
\begin{array}{clr}
\alpha \!\!\!\!\!\! &\times f_{low}(x)  & : x \leq x_0 \\
(1-\alpha) \!\!\!\!\!\! & \times f_{high}(x) & :  x_0 < x \leq1
\end{array}
\right.
\end{equation}

\noindent Here $x=T_\pi/T_\pi^{\rm max}$ with $T_\pi$ and $T_\pi^{\rm max}$ the $\pi$-meson kinetic energy and its maximum kinetic energy in the laboratory frame. The functional form of $f_{low}$ and $f_{high}$ are:
\begin{equation}
\begin{array}{l}
f_{low}(x)~ = N_{low}\times x \times \exp\left(-\beta\,x\right) \\
f_{high}(x) = N_{high}\times (1-x) \times \exp\left(-\gamma\,x\right),
\end{array}
\end{equation}

\noindent and the normalisation constants $N_{low}$ and $N_{high}$ are defined such that the integral of $f_{low}$ from 0 to $x_0$ is one and the integral of $f_{high}$ from $x_0$ to 1 is one and have the form:
\begin{equation}
\begin{array}{l}
N_{low}=\frac{\beta^2}{1-(1-\beta\,x_0)\exp\left(-\beta\,x_0\right)}\\
\\
N_{high}=\frac{\gamma^2}{\exp\left(-\gamma\right) - (1-\gamma(1-x_0))\exp\left(-\gamma\,x_0\right)}
\end{array}
\end{equation}

The derived functional form of $x_0$, $\alpha$, $\beta$ and $\gamma$ for $T_p\leq 10$~GeV are shown below: 
\begin{equation}
x_0 = \left\{
\begin{array}{ll}
 0.17  & : T_p < 1.5~{\rm GeV} \\
 0.10  & : 1.5 \leq T_p < 5~ {\rm GeV}\\
 0.08  & : 5 \leq T_p \leq 10~ {\rm GeV}
\end{array}
\right.
\end{equation}

\begin{equation}
\alpha = \left\{
\begin{array}{ll}
 0.42 - 0.1\,\theta_p  & : T_p < 1 ~{\rm GeV} \\
 0.36  & : 1 \leq T_p < 1.5 ~{\rm GeV}\\
 0.288 +\frac{S(18[\theta_p-1.85]) + 10\,S(1.7[\theta_p-4.1])}{40} & : 1.5 \leq T_p \leq 10 ~{\rm GeV}\\
\end{array}
\right.
\end{equation}

\begin{equation}
\beta = \left\{
\begin{array}{ll}
 101\,\theta_p^3 -230\,\theta_p^2 +170\,\theta_p -30  & : T_p \leq 1 ~{\rm GeV} \\
 12.3  & : 1 \leq T_p < 1.5 ~{\rm GeV}\\
 18.5 + 10\,S(13[\theta_p-3.4])  & : 1.5 \leq T_p \leq 10 ~{\rm GeV}\\
\end{array}
\right.
\end{equation}

\begin{equation}
\gamma= \left\{
\begin{array}{ll}
 2.5 - 2\,\theta_p  & : T_p \leq 1 ~{\rm GeV} \\
 0.68  & : 1 \leq T_p < 1.5 ~{\rm GeV}\\
 1.9\,\theta_p^{1/2} -1.1 & : 1.5 \leq T_p \leq 10 ~{\rm GeV}\\
\end{array}
\right.
\end{equation}

\noindent The $S(x)=\left[1+\exp(-x)\right]^{-1}$ is the sigmoid function and $\theta_p=T_p/m_p$ with $T_p$ and $m_p$ are the proton kinetic energy and mass, respectively.

\end{appendix}

\bibliographystyle{aa}
\bibliography{secondary}
\end{document}